\documentclass[conference]{IEEEtran}
\ifCLASSINFOpdf
\else
\fi
\usepackage{amsmath}
\usepackage{yfonts}
\usepackage{amsmath,amssymb,amsthm,xcolor,enumitem,stmaryrd,multicol}
\usepackage{tikz,bbm,mathrsfs}
\usepackage{nicefrac}
\usepackage{graphicx}
\usepackage[all]{xy}
\usepackage[appendix=inline]{apxproof}

\usepackage{centernot}
\newcommand{\srelmacro}{\mathrel{\vbox{\offinterlineskip\ialign{\hfil##\hfil\cr$\longrightarrow$\cr\noalign{\kern-.4ex}$\centernot\longleftarrow$\cr}  \vspace{-4pt}}}}

\newcommand{\ignore}[1]{}

\newcommand{\access}{\longrightarrow}
\newcommand{\biaccess}{\longleftrightarrow}

\newcommand{\revaccess}{\longleftarrow}

\newcommand{\rel}{\access}
\newcommand{\rrel}{\access^*}
\newcommand{\simrel}{\biaccess}
\newcommand{\srel}{\srelmacro}

\newcommand{\crel }{\access^*}

\newcommand{\pvar}{{\mathbb P}}
\newcommand{\free}{{\rm Free}}
\newcommand{\refbox}{{[*]}}
\newcommand{\refdia}{{\langle *\rangle}}
\newcommand{\upst}{{{\uparrow}}}
\newcommand{\rupst}{{{\uparrow}^*}}

\newcommand{\rdnst}{{\downarrow^*}}

\newcommand{\tangle}{\Diamond^\infty}
\newcommand{\reftangle}{\langle *\rangle^\infty}
\newcommand{\f}{\varphi}

\newcommand{\tuple}{\overline}

\newcommand{\bM}{{\mathcal M}}

 \newcommand{\david}[1]{}

\newcommand{\placeholder}{\cdot}
\newcommand{\dpt}{{\rm dpt}}

\newcommand{\ps}{\Diamond}

\usepackage{amssymb,graphicx}

\newcommand{\dlap}[1]{\makebox[0pt]{\hss#1\hss}}
\newcommand{\cd}{%
  \sbox0{$\lozenge$}%
  \usebox0\kern-.5\wd0\dlap{\raisebox{.1ex}{\scalebox{.7}[1]{$\cdot$}}}\kern.5\wd0%
}

\newcommand{\cb}{%
  \sbox0{$\Box$}%
  \usebox0\kern-.5\wd0\dlap{\raisebox{.2ex}{\scalebox{.7}[1]{$\cdot$}}}\kern.5\wd0%
}

\def\<{\left <}

\def\>{\right >}

\DeclareSymbolFont{AMSb}{U}{msb}{m}{n}
\DeclareMathSymbol{\N}{\mathbin}{AMSb}{"4E}
\DeclareMathSymbol{\Z}{\mathbin}{AMSb}{"5A}
\DeclareMathSymbol{\R}{\mathbin}{AMSb}{"52}
\DeclareMathSymbol{\Q}{\mathbin}{AMSb}{"51}
\DeclareMathSymbol{\I}{\mathbin}{AMSb}{"49}
\DeclareMathSymbol{\C}{\mathbin}{AMSb}{"43}

\newtheorem{thm}{Theorem}[section]
\newtheorem{remark}[thm]{Remark}
\newtheorem{proposition}[thm]{Proposition}
\newtheorem{definition}[thm]{Definition}
\newtheorem{lemma}[thm]{Lemma}
\newtheorem{prop}[thm]{Proposition}
\newtheorem{corollary}[thm]{Corollary}
\newtheorem{cor}[thm]{Corollary}
\newtheorem{exam}[thm]{Example}

\newtheoremrep{thmrep}[thm]{Theorem}
\newtheoremrep{remarkrep}[thm]{Remark}
\newtheoremrep{propositionrep}[thm]{Proposition}
\newtheoremrep{definitionrep}[thm]{Definition}
\newtheoremrep{lemmarep}[thm]{Lemma}
\newtheoremrep{proprep}[thm]{Proposition}
\newtheoremrep{corollaryrep}[thm]{Corollary}
\newtheoremrep{correp}[thm]{Corollary}
\newtheoremrep{examrep}[thm]{Example}

\begin{document}
\title{The Topological Mu-Calculus:\\
completeness and decidability}

\author{\IEEEauthorblockN{Alexandru Baltag}
\IEEEauthorblockA{University of Amsterdam\\
Email: thealexandrubaltag@gmail.com}
\and
\IEEEauthorblockN{Nick Bezhanishvili}
\IEEEauthorblockA{University of Amsterdam\\
Email: N.Bezhanishvili@uva.nl}
\and
\IEEEauthorblockN{David Fern\'andez-Duque}
\IEEEauthorblockA{Ghent University\\
Email: David.FernandezDuque@UGent.be}
}


%Telephone: (800) 555--1212, Fax: (888) 555--1212}



\IEEEoverridecommandlockouts
\IEEEpubid{\makebox[\columnwidth]{978-1-6654-4895-6/21/\$31.00~
\copyright2021 IEEE \hfill} \hspace{\columnsep}\makebox[\columnwidth]{ }}

\maketitle

\begin{abstract}
We study the topological $\mu$-calculus, based on both Cantor derivative and closure modalities, proving completeness, decidability and FMP over general topological spaces, as well as over $T_0$ and $T_D$ spaces. We also investigate relational $\mu$-calculus, providing general completeness results for all natural fragments of $\mu$-calculus over many different classes of relational frames. Unlike most other such proofs for $\mu$-calculus, ours is model-theoretic, making an innovative use of a known Modal Logic method (--the 'final' submodel of the canonical model), that has the twin advantages of great generality and essential simplicity.
\end{abstract}


% \fi
\IEEEpeerreviewmaketitle

\section{Introduction}

The modal $\mu$-calculus is one of the most powerful extensions of modal logic, of great use in computer science applications.
It is decidable, but very expressive, embedding many modal/temporal logics, such as $\sf PDL$, $\sf CTL$ and $\sf{CTL}^\ast$, that are widely applied in program specification and verification.

The completeness of the modal $\mu$-calculus was a difficult problem and remained open for many years.
Even today, there are very few completeness results for axiomatic systems for $\mu$-calculus with respect to standard classes of Kripke models (e.g., \cite{Waluk00,Santocanale08,SantVen10}, see also a more recent proof theoretic approach \cite{AL17}).
Prior to our work, there seemed to be no general model-theoretic method to establish completeness for various natural fragments of $\mu$-calculus over various classes of models.

An alternative interpretation of modal logic is based not on Kripke frames, but on topological spaces.
This semantics is in fact older and can be traced back to McKinsey and Tarski \cite{MT44}.
When the modal $\Diamond$ is interpreted as topological closure and the modal $\Box$ as topological interior, one obtains a semantics for the modal logic $\sf S4$ and its extensions, generalizing Kripke semantics over transitive, reflexive frames.
The logic of all topological spaces in this semantics is $\sf S4$.
We refer to \cite{vBB07} for an overview of the rich landscape of results on topological completeness of modal logics above $\sf S4$.

McKinsey and Tarski also suggested a second topological semantics, obtained by interpreting the modal $\Diamond$ as Cantor derivative.\footnote{Recall that the derivative $d(A)$ of a set $A$ consists of all limit  points of $A$.}
Esakia \cite{Esakia01, Esakia04} showed that the derivative logic of all topological spaces is the modal logic $\mathsf{wK4} = \mathsf{K} + (\Diamond \Diamond p \to \Diamond p\vee p$).  This is also the modal logic of all {\em weakly transitive} frames, i.e.~those for which the reflexive closure of the accessibility relation is transitive.
It is well-known that the modal logic of transitive frames is $\mathsf{K4}$ \cite{BdRV01, CZ97}, which moreover corresponds to a natural class of topological spaces denoted $T_D$. Another natural class are $T_0$ spaces, whose modal logic is also finitely axiomatizable; we discuss $T_0$ spaces and $T_D$ spaces in the context of modal logic in Section \ref{secMu}.
Modal logics axiomatizing well-known classes of spaces also include the G\"odel-L\"ob logic $\sf GL$, which is complete with respect to the class of scattered spaces \cite{Esakia81, BeklGab14}.

Topological structures are of great interest to Computer Science. As noticed by Vickers \cite{Vick89} and Abramsky \cite{Abr91}, the notion of observability and its logic require a topological setting. Abstract notions of computability also involve topological structures, and a famous example is Scott topology. More recently, developments in Formal Learning Theory \cite{Brecht,Solvability}, Distributed Computing \cite{TopoComp} and Epistemic Logic in Multi-Agent Systems \cite{RausbaumDEL,Ayb17,BBOS16}, have taken a topological turn. In particular, recent epistemic work \cite{BBOS16, Ayb17} on modelling and reasoning about evidence and knowability uses topological structures. Research on spatial reasoning, in both topological and metric incarnations, is also of significant interest for AI. The addition of fixed point operators allows us to reason about non-trivial properties of topological spaces: for example, the well-known Cantor-Bendixon theorem states that any topological space has a {\em perfect core}, i.e. a maximal subset equal to its own derivative.
The perfect core is not modally definable (in terms of derivative or closure modalities), but it is definable in the $\mu$-calculus with the derivational semantics. Parikh \cite{Par92} showed the relevance of Cantor derivative and the perfect core for multi-agent epistemic puzzles and applications.\footnote{In on-going work, we show that the perfect core and its logic have deep connections with the topic of learnability from observations, as well as with epistemic paradoxes, such as the Surprise Examination.}

\medskip

Our main aim in this paper is to investigate the \emph{topological} $\mu$-calculus based on the Cantor derivative modality, as well as
its weaker version based on the closure modality. As a secondary aim, we explore (various fragments of) the \emph{relational} $\mu$-calculus, on (various classes of) weakly transitive frames.
As such, our contribution in this paper is two-fold.
First, we develop a new model-theoretic method of proving completeness for systems of $\mu$-calculus over weakly transitive frames.
This method applies to a wide range of logics, including many well-known ones. Concretely, we show that if a modal logic $\Lambda$ is a canonical cofinal subframe logic, then its modal $\mu$-variant, obtained by adding the fixed-point axiom and induction rule, is Kripke complete
and enjoys the finite model property.
This implies that the modal $\mu$-variants of the well known modal logics $\sf{wK4},  \sf{wKT_0},  \sf{K4}, \sf{K4D}, \sf{K4.1}, \sf{K4.2}, \sf{K4.3}, \sf{S4}, \sf{S4.1},
\sf{S4.2}$, and $\sf{S4.3}$ have the FMP\footnote{In fact, there are continuum-many such logics \cite{CZ97}, so our results apply to uncountably many classes of frames.} and are decidable.  Second, we show that the
derivational $\mu$-calculus is completely axiomatized on all topological spaces, all $T_0$ spaces, and all $T_D$ spaces, by the $\mu$-variants of the logics $\sf wK4$, $\sf wK4T_0$, and $\sf K4$, respectively. We also give a new proof of the known fact that the weaker $\mu$-calculus based on topological closure
is completely axiomatized by the $\mu$-variant of the modal logic $\sf S4$.

Our model-theoretic proof is based on restricting the canonical model to the set of {\em final theories}, i.e. theories which satisfy a natural maximality condition.
A similar construction has been employed by Fine \cite{Fin85} to prove FMP for subframe logics over $\sf K4$ (without fixed point operators).
Zakharyaschev \cite{Zak96} generalized this to show FMP for cofinal subframe logics over $\sf K4$, and \cite{BGJ11} extended this result to cofinal subframe logics over $\sf wK4$.
Our Kripke-completeness results apply to the $\mu$-variants of the same class of logics.
The crucial new insight is that the truth lemma extends to the full $\mu$-calculus over the set of final theories, despite not doing so for the full canonical model. Topological completeness follows then from more or less standard constructions and the observation that the logics of the classes of all topological spaces, all $T_0$ spaces, and all $T_D$ spaces are all subframe logics.

These results are new, with one proviso concerning $T_D$ spaces.
First, note that the transitive closure of a binary relation is definable in standard relational $\mu$-calculus (on arbitrary frames).
Thus, FMP for $\mu$-calculus over transitive frames follows immediately from Kozen's proof of FMP for general $\mu$-calculus \cite{Kozen}. Second, Goldblatt and Hodkinson \cite{GH18} have completely axiomatized a modal logic (with the so-called \emph{tangled derivative} modality), that is co-expressive with derivational $\mu$-calculus over $T_D$ spaces, by a result of Dawar and Otto \cite{DO09}. But, as explained in Section \ref{Conclusion}, even in the $T_D$ case, our work has the added benefit of providing a direct proof of completeness/decidability of full $\mu$-calculus over these spaces, without relying on the corresponding results for standard $\mu$-calculus.
Moreover, dropping the $T_D$ assumptions drastically changes the behavior of the $\mu$-calculus in at least two important ways. Weakly transitive closure does \emph{not} seem to be definable in $\mu$-calculus, and so decidability over arbitrary (as well as $T_0$) spaces does \emph{not} follow from any known results. Finally, as shown in Section \ref{secTan}, the above-mentioned co-expressivity result fails on arbitrary (or $T_0$) spaces: $\mu$-calculus on such spaces does \emph{not} collapse to its tangled fragment.
For this reason, we work here with the full language of $\mu$-calculus.

\bigskip

\par\noindent\textbf{The structure of this paper} In Section II we introduce derivative spaces, a general notion subsuming both topological spaces and weakly-transitive frames. Section III defines $\mu$-calculus over such spaces and states our main completeness result. In Section IV, we show that the tangled fragment is not expressively complete in this setting. Section V investigates truth-preserving maps and relations between derivative spaces. Section VI presents the stepping stones of the main completeness proof. Section VII generalizes this to an infinite class of fixed-point logics, while Section VIII extends it to $T_0$ and $T_D$ spaces. We end in Section IX with some concluding remarks and a comparison with related work. All the proof details are in the Appendix.

\section{Derivative spaces}\label{DerivSpace}

Although our primary focus in this paper is the derivational $\mu$-calculus on topological spaces, for technical reasons it is useful to consider a slightly more general class of structures.

\begin{definition}
A \emph{derivative space} is a pair $(\mathcal{X},d)$, where $\mathcal{X}$ is a set of `points', and $d:\mathcal{P}(\mathcal{X})\to \mathcal{P}(\mathcal{X})$ is an operator on subsets of $\mathcal{X}$, satisfying the following properties, for all $X,Y\subseteq \mathcal{X}$:
\begin{itemize}
\item $d(\varnothing)=\varnothing$;
\item $d(X\cup Y)=d(X)\cup d(Y)$;
\item $d(d(X))\subseteq X\cup d(X)$.
\end{itemize}
The conjunction of the first two conditions above is known as \emph{normality}, while the third condition is known as \emph{weak idempotence}.
\end{definition}

The notion of derivative space is the concrete set-theoretic instantiation of the more abstract concept of \emph{derivative algebra}, introduced by Esakia \cite{Esakia04} (as a generalization of a notion with the same name introduced by McKinsey and Tarski \cite{MT44}).
%If we add the axiom $X\subseteq d(X)$ (and, {\em a fortioiri,}
% \emph{transitivity:}\david{ALEXANDRU: To be consistent, you'd need to call this idempotence.}
%$d(d(X))\subseteq d(X)$), we obtain the {\em Kuratowski axioms} for {\em closure spaces.}
%In this case, $d$ is usually denoted $c$ (for `closure').
%It is well-known that closure spaces are in one-one correspondence with topological spaces.
%\smallskip

%{\bf Nick: Maybe examples shouldn't be italicized? Also if we want to cut text maybe we could cut one of them or merge two? }

\begin{exam}[topological closure spaces]
A special case of derivative spaces is given by \emph{closure spaces}:
these are derivative spaces $(\mathcal{X}, c)$ that additionally satisfy $X\subseteq c(X)$ (and, {\em a fortiori,} $c(c(X))\subseteq c(X)$).
These strengthened conditions are known as the {\em Kuratowski axioms,} that define topological spaces in terms of their closure operator.\footnote{Given a closure space, let  $X\subseteq\mathcal X$ be {\em closed} whenever $X=c(X)$, and {\em open} whenever its complement is closed. This gives us the more common definition of topology as a family of open or closed sets. So closure spaces are exactly the same notion as topological spaces.} When considered as a special case of derivative spaces, with $d(X):=c(X)$ given by topological closure, topological spaces will be called \emph{topological closure spaces}.
\end{exam}

\begin{exam}[topological derivative spaces]
Our main example of derivative spaces in this paper are structures
$(\mathcal{X},d)$, based on an underlying topological (closure) space $(\mathcal{X},c)$ (satisfying the Kuratowski axioms), but with the derivative operator given by the so-called \emph{Cantor derivative}, i.e. by taking $d(X)$ to be the \emph{set of limit points} of $X$:
\begin{align*}
d(X) & :=\, \{y\in \mathcal{X}: y\in c(X-\{y\})\} \\
&= \{y\in \mathcal{X}:\forall U\in \mathcal{N}(y)\, X\cap (U-\{y\})\not=\emptyset\},
\end{align*}
where $\mathcal{N}(y)$ is the family of (open) neighborhoods of $y$ in the space $(\mathcal{X},c)$.
It is easy to see that  $(\mathcal{X},d)$ is a derivative space, which we'll refer to as a \emph{topological derivative space}.
The closure operator can be recovered as $c(X) = X\cup d(X)$.
\end{exam}
%Obviously, these are nothing but \emph{topological spaces} $(\mathcal{X},c)$, but in which the `derivative' operator is given by topological \emph{closure} $d_c(X):=c(X)$.

%Our primary example is a topological space $(\mathcal{X},c)$, say given in terms of a closure operator $c:\mathcal{P}(\mathcal{X})\to \mathcal{P}(\mathcal{X})$ satisfying the Kuratowski closure axioms
%\footnote{These are: $c(\varnothing)=\varnothing$; $X\subseteq c(X)$; $c(X\cup Y)=c(X)\cup c(Y)$; $c(X)=c(c(X))$.},

So, every topological space gives rise to a derivative space in at least two different ways (as a closure space, and as a topological derivative space), though we are mostly interested in the second one.
The converse is also true:

\smallskip

\par\noindent\textbf{Closure and interior in derivative spaces}
Given a derivative space  $(X,d)$, we define the \emph{closure} and \emph{interior} operators $c, i:\mathcal{P}(\mathcal X)\to \mathcal{P}(\mathcal X)$, by putting
$$c(X):=X\cup d(X), \,\,\,\,\,\, i(X)\, := \, \mathcal{X}- c(\mathcal{X}-X).$$
It is easy to see that these satisfy all the Kuratowski axioms.

This means that every derivative space induces a topological space. Moreover, in a \emph{topological} derivative space (with Cantor derivative over some topological space), the induced closure operator (as defined above) coincides with the underlying topological closure. But in general, this matching does not work the other way around: given an arbitrary derivative space, its derivative does \emph{not} necessarily coincide with the Cantor derivative in the induced topology (given by the above-defined closure operator). It follows that \emph{not every derivative space is a topological derivative space}. A counterexample is given by the next special case.
% The converse is also true:
%
%\par\noindent\textbf{Closure and interior in derivative spaces}
%Given a derivative space  $(X,d)$, we define the \emph{closure} and \emph{interior} operators $c, i:\mathcal{P}(\mathcal X)\to \mathcal{P}(\mathcal X)$, by putting
%$$c(X):=X\cup d(X), \,\,\,\,\,\, i(X)\, := \, \mathcal{X}- c(\mathcal{X}-X).$$
%It is easy to see that these satisfy all the Kuratowski axioms.
%
%
\begin{exam}[weakly transitive Kripke frames] A \emph{weakly transitive frame} (or {\em $\sf wK4$ frame}) is a Kripke structure $(W,\rel)$ (also known as a `transition system'), consisting of a set of `states' (or `possible worlds') $W$, together with a binary relation ${\rel}\subseteq W\times W$ (known as an `accessibility' or `transition' relation), assumed to be \emph{weakly transitive}: i.e., for all states $w,s,t\in W$, if $w \rel s\rel t$ then either $w=t$ or $w \rel t$. We denote by $\rrel$ the reflexive closure ${\rm Id}\cup {\rel}$ of $\rel$, which (due to weak transitivity) coincides with its transitive-reflexive closure ${\rm Id}\cup \bigcup_{n\geq 1} \access^n$.

We also denote by $\srel$ the strict part of $\rel$, i.e. $w\srel v$ if $w\rel v\centernot{\rel} w$; and write $w\simrel v $ if $w\rel v\rel w$ and $w\simrel^* v $ if $w\simrel v $ or $w=v$.
For any state $w\in W$, we put $w{{\uparrow}}:=\{s\in W: w\rel s\}$ for the set of its successors, and also put $w\rupst:=\{s\in S: w\rrel s\}=\{w\}\cup w{{\uparrow}}$; more generally, for any set $X\subseteq W$, we put $X{{\uparrow}}:=\{s\in W: x\rel s \mbox{ for some } x\in X\}=\bigcup_{x\in X} x{{\uparrow}}$, and similarly put $X\rupst:=\{s\in W: x\rrel s \mbox{ for some } x\in X\}=X\cup X{{\uparrow}}$. By applying the same definitions to the converse $\revaccess$, we obtain the corresponding notions of down-closure $w{\downarrow}$, $w{\rdnst}$, $X{\downarrow}$, $X{\rdnst}$.

It is easy to see that every weakly transitive frame gives rise to a derivative space $(\mathcal{X},d_{\rel})$, obtained by taking $\mathcal{X}:=W$, and taking the derivative $d_{\rel}$ to be usual modal `Diamond' operator:
\begin{align*}
d_{\rel}(X) & :=\,  X{\downarrow}=\{w\in W: X\cap {w{{{\uparrow}}}} \neq \varnothing\}\\
&=\{w\in W: \exists s \, w\rel s \in X\}.
\end{align*}
Moreover, the induced closure $c_{\rel}(X)$ (as defined above in arbitrary derivative spaces) is given by
$c_{\rel}(X)\, = \, X{\rdnst}.$
\end{exam}

In general, weakly transitive frames are \emph{not} topological derivative spaces. But the intersection of the two classes is of independent interest, as shown by the next two examples:

\begin{exam}[Alexandroff closure spaces as $S4$ Kripke frames]
A topological space $(\mathcal{X},c)$ is \emph{Alexandroff} if its closure operator distributes over arbitrary unions: $c(\bigcup_i X_i)=\bigcup_i c(X_i)$.
%(Equivalent conditions: the family of closed sets is closed under arbitrary unions; the family of open sets is %closed under arbitrary intersections; every point has a smallest neighborhood).
Given $x,y\in \mathcal X$, define $x\rel y$ if $x\in c\{y\}$.
Then, it is not hard to check that if $\mathcal X$ is Alexandroff, then $\rel$ is a reflexive-transitive relation, i.e.  $(\mathcal{X},\rel)$ is an $S4$ Kripke frame, and moreover the relational derivative coincides in this case with the topological closure: $d_{\rel}=c$. As it is well-known, the converse also holds: every $S4$ frame $(\mathcal{X},\rel)$ gives rise to an Alexandroff closure space, by putting $c_{\rel}(X):=X\downarrow=X\downarrow^*$ for the closure/derivative operator. This time, the equivalence is complete: starting from either side, and applying successively these two transformations, we obtain the original structure. So \emph{Alexandroff topological closure spaces are essentially the same as $S4$ Kripke frames.}
\end{exam}

\begin{exam}[Alexandroff derivative spaces as irreflexive $\sf wK4$ frames]\label{AlexDerivative}
Another way to convert an Alexandroff space $(\mathcal{X},c)$ into a relational structure is to define $x\rel y$ if $x\in d\{y\}=c\{y\} - \{y\}$, for all $x,y\in \mathcal{X}$. Then $\rel$ is weakly transitive and irreflexive, and the relational derivative $d_{\rel}$ coincides in this case with the Cantor derivative induced by $c$. Conversely, every irreflexive $\sf wK4$ frame $(\mathcal{X},\rel)$ gives rise to an Alexandroff derivative space $(\mathcal{X},d)$, by putting $c_{\rel}(X):=X\downarrow^*$ for the topological closure, and taking $d$ to be induced Cantor derivative in the resulting topology (for which one can check that $d(X)=X{\downarrow}$). Once again, the equivalence is complete: by applying successively these transformations,
we obtain the original structure. So \emph{Alexandroff topological derivative spaces are essentially the same as irreflexive $\sf wK4$ frames.}
\end{exam}

%Finally, there is another intersection, of the class of (topological) closure spaces and the one of weakly transitive frames: these are the reflexive-transitive frames.

%\begin{exam}[Alexandroff closure spaces as $\sf S4$ frames]
%Take an Alexandroff space $(\mathcal{X},c)$, in which now the `derivative' is given by the closure $d_c(X):=c(X)$. It is well-known that such Alexandroff closure spaces are essentially the same as \emph{reflexive-transitive frames} (also known as
%\emph{$\sf S4$ frames}): relational structures $(W,\rel)$, whose binary relation $\rel\subseteq W\times W$ is both reflexive and transitive. Note that in such frames we have ${X\rupst}={X{{\uparrow}}}$ and${X\rdnst}={X{\downarrow}}$, for all sets $X$. The correspondence is a special case of the one above: given an Alexandroff space $\mathcal{X}$ with closure $c$, we get an $S4$ frame by putting $x \rel y$ iff $y\in d_c(x)=c(x)$, for all $x,y\in \mathcal{X}$; and vice-versa, given a reflexive-transitive frame $(W,\rel)$, we can convert it to an Alexandroff space, by putting $c(X):={X\rdnst}={X{\downarrow}}$.
%\end{exam}

\medskip

\par\noindent\textbf{D-neighborhoods} For every point $x\in \mathcal{X}$ in a derivative space $(\mathcal{X}, d)$, we can define the family of $d$-neighborhoods of $x$:
$$\mathcal{N}_d(x)\,:=\, \{X\subseteq \mathcal{X}: x\not\in d(\mathcal{X}-X)\}$$
Note that, in general, d-neighborhoods are \emph{not} neighborhoods of $x$ in the topology given by the closure $c(X)$ induced by $d$. In fact, in a topological derivative space (where derivative means Cantor derivative), a d-neighborhood $X\in \mathcal{N}_d(x)$ is just a `punctured neighborhood' of $x$, i.e. a set with the property that $U-\{x\}\subseteq X$ for some open neighborhood $U\ni x$. On the other hand, in a topological closure space
(where the `derivative' is just the topological closure), d-neighborhoods coincide with standard topological neighborhoods. Finally, in a weakly transitive frame $(W,\rel)$, a set $X\subseteq W$ is a d-neighborhood of a state $x\in W$ iff $x{\uparrow} \subseteq X$.

We can now characterize the derivative in terms of d-neighborhoods, in a way that generalizes the definition of Cantor derivative in topological spaces:

\begin{lemma}\label{d-neighb}
For every set $X\subseteq \mathcal{X}$ in a derivative space $(\mathcal{X},d)$, we have
$$d(X)=\{y\in \mathcal{X}: \forall U\in \mathcal{N}_d(y)\,\, U\cap X\not=\varnothing\}.$$
\end{lemma}

This leads to an equivalent presentation of derivative spaces as a special case of monotonic neighborhood structures \cite{Pac17}: a \emph{neighborhood derivative space} is a pair $(\mathcal{X}, \mathcal{N})$, where $\mathcal{X}$ is a set of points, and $\mathcal{N}: \mathcal{X}\to \mathcal{P}(\mathcal{P}(\mathcal{X}))$ is a map that assigns to each point $x\in \mathcal{X}$ a family $\mathcal{N}(x)\subseteq \mathcal{P}(\mathcal{X})$ of `neighborhoods' of $x$, satisfying the following conditions

\begin{enumerate}
\item $\mathcal{X}\in \mathcal{N}(x)$;
\item if $X\in \mathcal{N}(x)$ and $X\subseteq Y$, then $Y\in \mathcal{N}(x)$;
\item if $X,Y\in \mathcal{N}(x)$, then $X\cap Y\in \mathcal{N}(x)$;
\item if $x\in X\in \mathcal{N}(x)$, then $\{y\in \mathcal{X}: X\in \mathcal{N}(y)\}\in \mathcal{N}(x)$.
\end{enumerate}

Each derivative space $(\mathcal{X}, d)$ gives rise to a neighborhood derivative space by taking $$\mathcal{N}(x)\, :=\, \mathcal{N}_d(x) = \{X\subseteq \mathcal{X}: x\not\in d(\mathcal{X}-X)\}$$
to be the set of all $d$-neighborhoods. Conversely, every neighborhood derivative space $(\mathcal{X}, \mathcal{N})$ gives rise to a derivative space, via the following generalization of Cantor derivative:
$$d(X)\, :=\, \{y\in \mathcal{X}: \forall U\in \mathcal{N}(y)\, U\cap X\not=\varnothing\}.$$
This is a full equivalence between derivative spaces and neighborhood derivative spaces: starting from either side, and applying
the above two transformations, we obtain the original structure.

\section{Mu calculus on derivative spaces: main results}\label{secMu}

For reasons having to do with our intended applications, as well as to simplify some proof details, in this paper we take the \emph{greatest fixed point} operator $\nu x. \f$ as primitive, and define the least fixed point $\mu x. \f$ as an abbreviation.\footnote{This setting is of course equivalent to the more standard presentation, that takes $\mu x. \f$ as primitive.}

\medskip

\par\noindent\textbf{Syntax}: Let $\pvar$ be a set of \emph{propositional variables}. We recursively define the set ${\mathcal L}_\mu$ of \emph{formulas}, together with a map $\free:L_\mu\to {\mathcal P}(\pvar)$, associating to each formula $\varphi\in {\mathcal L}_\mu$ its set of \emph{free variables} $\free(\varphi)\subseteq \pvar$. The definition is by simultaneous recursion, with formulas $\f\in {\mathcal L}_\mu$ given by
$$
\begin{array}{ccc cc cc ccc cc cc cc cc cc ccc ccc}
\varphi :: =  & \top &|& x &|& \neg \f  &|& \varphi\wedge \varphi &|& \Diamond \varphi &|& \nu x.\f
\end{array}
$$
where: $x\in \pvar$; in the construct $\f\wedge \f'$, no variables occur free in $\f$ and bound in $\f'$, or vice versa; and in the construct $\nu x. \f$, formula $\f$ is \emph{positive in $x$} (i.e. whenever $x$ occurs in $\f$, we have that $x\in \free(\f)$ and $x$ occurs only in the scope of an even number of negations). The set $\free(\f)$ of free variables of a formula $\f$ is simultaneously defined by recursion:
\begin{align*}
\free(\top)  :=  \varnothing,& \ \ \ \free(x):=\{x\}, \\
\free(\f\wedge \f')& :=\free(\f)\cup \free(\f'),\\
\free(\neg\f) &= \free(\Diamond\f):=\free(\f) ,\\
\free(\nu x.\f) &:= \free(\f)-\{x\}.
\end{align*}

A variable is \emph{bound} in $\f$ if it occurs in $\f$ but is not in $\free(\f)$.
For any set of variables $P\subseteq \pvar$, we denote by $\mathcal{L}_\mu^P$ the set of all formulas $\f\in\mathcal{L}_\mu$ having $\free(\f)\subseteq P$.
Note in particular that $\mathcal L_\mu = \mathcal L_\mu^\pvar$.

We use the notation $\overline{x}=(x_1, \ldots, x_n)$ to denote finite strings of variables $x_1, \ldots, x_n\in \pvar$, and denote by $\lambda$ the empty string. When we want to make explicit some of the free variables, we write $\f(\overline{x})$
for a formula in which all variables in the string $\overline{x}$ are free (if occurring at all).

\medskip

\par\noindent\textbf{Subformulas} The \emph{subformula relation}   $\sqsubset$ is the smallest transitive relation on formulas satisfying the following properties: $\f\sqsubset (\neg\f),(\Diamond\f),  (\nu x. \f)$, and $\f\sqsubset (\f\wedge \f'), (\f'\wedge\f)$. The set ${\rm Sub}(\f)$ of all (improper) subformulas of $\f$ is defined as ${\rm Sub}(\f):=\{\f': \f'\sqsubset \f\} \cup\{\f\}$.

\medskip

\par\noindent\textbf{Semantics}. An \emph{atomic valuation} on a derivative space $(\mathcal{X},d)$ is a map
$\|\placeholder\|: \pvar\to {\mathcal P}(\mathcal{X})$ associating to each propositional atom $x\in \pvar$ some set of states $\|x\|\subseteq \mathcal{X}$. For each atomic valuation $\|\placeholder\|: \pvar\to {\mathcal P}(\mathcal{X})$, tuple $\tuple{x}=(x_1,\ldots, x_n)$ of variables and corresponding tuple $\tuple{X}=(X_1, \ldots X_n)$ of sets of points $X_i\subseteq \mathcal{X}$, we denote by $\|\placeholder\|_{\tuple{x}:=\tuple{X}}$ the valuation that assigns to each variable $x_i$ the set $X_i$ and agrees with the original valuation $\|\placeholder\|$ on all the other atoms.

A \textit{derivative model} $\bM=(\mathcal{X},d , \|\placeholder\|)$ consists of a derivative space $(\mathcal{X},d)$, together with an atomic valuation $\|\placeholder\|: \pvar\to {\mathcal P}(\mathcal{X})$.
The semantics is given by extending the atomic valuation to a map $\|\placeholder\|:\mathcal{L}_\mu\to \mathcal{P}(\mathcal{X})$, which we call \emph{extended valuation} (and for which we use the same notation $\|\placeholder\|$ as for the corresponding atomic valuation). The definition of the extended valuation is by recursion on subformulas: for propositional variables this is already given by the atomic valuation map of the model $\bM$, while in the rest we put
$$\|\top\|=\mathcal{X}, \,\,\,
\|\neg \f\|=\mathcal{X}-\|\f\|, \,\,\, \|\f\wedge \f'\|=\|\f\|\cap\|\f'\|,$$
$$\|\Diamond\f\|=d(\|\f\|), \,\,\,\,\,\,\,\,\, \|\nu x.\f\|= \bigcup \{X\subseteq \mathcal{X}: X\subseteq \|\f\|_{x:=X}\}.$$
For formulas $\f=\f(\tuple{x})$ and corresponding tuples of sets $\tuple{X}$, we will sometimes write $\|\f(\tuple{X})\|$ instead of $\|\f\|_{\tuple{x}:=\tuple{X}}$, in order to avoid subscript overload. With this notation, e.g., the clause for $\nu x$ becomes:
$\|\nu x.\f\|= \bigcup \{X\subseteq \mathcal{X}: X\subseteq \|\f(X)\|\}$.

Whenever $x\in \|\f\| $ for some point $x\in \mathcal{X}$, we also write $x\models_{\bM} \f$, and say that $\f$ is \emph{true} (or \emph{satisfied}) at point $x$ in the model $\bM$.
As usual, when the model is understood, we skip the subscript, writing $x\models\f$.
Conversely, we may write $\|\placeholder\|_\bM$ instead of $\|\placeholder\|$ when we wish to specify the relevant model.
We say that $\f$ is \emph{valid on the model $\bM$} if $\|\f\|_{\bM}=\mathcal{X}$, i.e. $\f $ is true at all points of $\bM$; similarly, $\f$ is \emph{satisfied on the model $\bM$} if  $\|\f\|_{\bM}\not=\varnothing$. By abstracting away from the specific valuation, we say that $\f$ is \emph{valid on the space} $(\mathcal{X},d)$ if for every valuation $\|\placeholder\|$ on $\mathcal{X}$, $\f$ is valid on the model $(\mathcal{X}, d, \|\placeholder\|)$; and $\f$ is \emph{satisfied on the space} $(\mathcal{X},d)$ if there exists a valuation $\|\placeholder\|$ on $\mathcal{X}$, s.t. $\f$ is satisfied on the model $(\mathcal{X}, d, \|\placeholder\|)$. Finally, $\f$ is \emph{valid (on a class $\mathfrak C$ of derivative models, or of derivative spaces)} if it is valid on all models/spaces (in the class $\mathfrak C$).
\smallskip

Note that in the special case of $\sf wK4$ relational models $(W,\rel)$, the above semantics of $\Diamond$ coincides with the standard Kripke semantics. As a consequence, on relational frames our semantics for $\mu$-calculus coincides with the standard one.

\medskip

\par\noindent\textbf{Abbreviations}: We have the usual abbreviations $\bot$, $\f\vee \psi$, $\f\Rightarrow\psi$, $\f\Leftrightarrow \psi$, $\Box\f$. The \emph{least fixed-point} formula $\mu x. \f(x,\tuple{y})$ can be defined as $\neg \nu x \neg \f (\neg x, \tuple{y})$. Finally, we define \emph{closure} and \emph{interior} modalities, as well as
\emph{tangled derivative}  $\tangle \Gamma$ and \emph{tangled closure} $\reftangle \Gamma$ (for finite sets of formulas $\Gamma$),
 with the \emph{perfect core} modality $\Diamond^\infty\f$ as a special case:
$$
\refdia\varphi : = \varphi \vee \Diamond \varphi,   \,\,\,\, \,\,\,\, \,\,\,\, \,\,\,\, \refbox\varphi  : = \varphi\wedge \Box\varphi,
$$
$$
\tangle \Gamma : =  \nu x. \bigwedge_{\gamma\in \Gamma} \Diamond (x \wedge \gamma), \,\,\,\, \,\,\,\, \,\,\,\,
\reftangle \Gamma : =  \nu x. \bigwedge_{\gamma\in \Gamma} \refdia (x \wedge \gamma),$$
$$\,\,\,\,  \, \, \Diamond^\infty\f  : =  \tangle \{\f\}. \,\,\,\, \,\,$$
%\ignore{
Note that the definitions of $\refdia\varphi$ and $\refbox\f$ do not use any fixed points. But, to justify these notations, one can easily check that, in the special case of weakly transitive frames, $\refbox$ and $\refdia$ are the standard Kripke modalities for the reflexive-transitive closure $\rel^*$ of the accessibility relation (which, as already mentioned, coincides on these frames with its reflexive closure). More generally,
in derivative spaces, $\refbox\varphi$ is equivalent to $\nu x. (\f \wedge \Box x)$,
while $\refdia\varphi$ is equivalent to $\neg\refbox\neg\f$ and thus to $\mu x. (\f \vee \Diamond x)$.
In fact, $\|\refbox\varphi\|$ and $\|\refdia\f\|$ coincide with the \emph{interior} $i(\|\f\|)$ and respectively the \emph{closure} $c(\|\f\|)$, as defined in derivative spaces.
In particular, in the case of \emph{topological} derivative spaces (where $d$ is Cantor derivative), these coincide with the underlying topological interior and closure operators. As for $\tangle \Gamma$ and $\reftangle \Gamma$, they are variants of the \emph{tangle modality} introduced in a relational setting by Dawar and Otto \cite{DO09}, who showed that $\mu$-calculus over transitive frames collapses to tangle logic based on $\tangle \Gamma$. Their topological interpretations were developed by Fernandez-Duque \cite{F-D11},
who distinguished between the \emph{tangled derivative}  $\tangle \Gamma$ and tangled closure $\reftangle \Gamma$, and axiomatized the logic of tangled closure. More recently, Goldblatt and Hodkinson \cite{GH18} axiomatized the logic of tangled derivative $\tangle \Gamma$ over transitive frames, and showed that it is equivalent to the logic over $T_D$ spaces. Finally, the perfect core modality $\Diamond^\infty\f$ is a special case of tangle, that captures Cantor's \emph{perfect core}: the largest subset of the state space that is equal to its own Cantor derivative.

\medskip

\par\noindent\textbf{Substitution and natural sublanguages} Given a formula $\f=\f(\overline{x})$ and a tuple of formulas $\overline{\theta}=(\theta_1, \ldots, \theta_n)$, we denote by $\f(\overline{\theta})$ the result of substituting every variable in $\overline{x}$ by the corresponding formula in $\tuple{\theta}$. Note that we have $$\|\f(\tuple{\theta})\|=\|\f(\tuple{\|\theta\|})\|$$
(where on the right hand we used an instance of the above-mentioned simplified notation $\|\f(\tuple{X})\|$ for $\|\f\|_{\tuple{x}:=\tuple{X}}$).
A {\em natural sublanguage} of ${\mathcal L}_\mu$ is any set $\mathcal L\subseteq {\mathcal L}_\mu$ which contains $\top$, is closed under substitution, and such that if $\varphi,\psi\in {\mathcal L}_\mu$ then also $\neg\varphi,\varphi\wedge\psi,\ps\psi\in \mathcal L$.
The basic modal language is a natural sublanguage and will be denoted $\mathcal L_\ps$.

\medskip

 The following characterization of  $\|\nu y. \f\|$ is also well-known in the literature:

\begin{prop}\label{Mono}
Let $(\mathcal X,d,\|\placeholder\|)$ be any derivative model and $\f=\f(y,\overline{x})$ be a formula that is positive in $y$. Then we have the following:
\begin{enumerate}
\item  the unary operator $Y\mapsto \|\f (Y, \overline{X})\|$ is \emph{monotonic}: if $Y\subseteq Y'$ then $\|\f(Y,\overline{X})\|\subseteq \|\f(Y',\overline{X})\|$;

\item $\|\nu y. \f(y,\overline{X})\|$ is the \emph{greatest fixed point} of the operator $Y\mapsto \|\f (Y, \overline{X})\|$,  i.e. the largest set $Y\subseteq \mathcal{X}$ s.t. $Y= \|\f (Y, \overline{X})\|$;

 \item  $\|\nu y. \f(y,\overline{X})\|=\bigcap_{\alpha\in {\sf On}} \f_y^\alpha(\tuple{X}),$
  where ${\sf On}$ is the class of ordinals and the transfinite sequence of sets $\f_y^\alpha(\tuple{X})\subseteq \mathcal{X}$ is defined by ordinal recursion:
     $\f_y^\alpha(\tuple{X})=\bigcap_{\beta<\alpha}  \|\f(\f_y^\beta(\tuple{X}), \tuple{X})\|$
     (and so in particular $\f_y^0(\tuple{X})=\mathcal X$).
\end{enumerate}
\end{prop}

\proof Well known (and easy to check).
\endproof

\medskip

\begin{definition}[$\mu\text{-}\mathsf{wK4}$]\label{defwK4mu}
We define the logic $\mu\text{-}\mathsf{wK4}$ to be the least set of formulas of $\mathcal L_\mu $ containing the following axioms and closed under the following rules (for all formulas $\f,\psi$, and formulas $\theta=\theta(x)$ that are positive in $x$):

\begin{itemize}

\item All the instances of the \textbf{Axioms and Rules of Propositional Logic}.

\item \textbf{Necessitation Rule}: From $\f$, infer $\Box \f$.

\item \textbf{Distribution Axiom} (=Kripke's Axiom $K$):
\\  $\Box(\f\Rightarrow\psi)\Rightarrow (\Box\f \Rightarrow \Box\psi)$.

\item \textbf{Weak Transitivity}:
$\Diamond \Diamond \f \, \Rightarrow \, (\f\vee \Diamond\f)$.

\item \textbf{Fixed Point Axiom}:
$\nu x. \theta \, \Rightarrow \, \theta(\nu x. \theta)$.

\item \textbf{Induction Rule}:  From $\f \, \Rightarrow \, \theta(\f)$,
infer $\f\, \Rightarrow \, \nu x. \theta$.
\end{itemize}
\end{definition}

We will also be interested in variants of $\mu\text{-}\mathsf{wK4}$.
If $\Lambda$ is any normal modal logic over $\mathcal L_\ps$ (in the sense of \cite{BdRV01}) that extends $\mathsf{wK4}$, then $\mu\text{-}\Lambda$ is the extension of $\mu\text{-}\mathsf{wK4}$ with all axioms of $\Lambda$, closed under uniform substitution with arbitrary formulas in $\mathcal L_\mu$.
If $\mathcal L$ is a natural sublanguage of $\mathcal L_\mu$, then $\mu\text{-}\Lambda^\mathcal L$ is the restriction of $\mu\text{-}\Lambda$ to $\mathcal L$, in the sense that all axioms and rules may only be applied when all formulas belong to $\mathcal L$.

\begin{prop}[Soundness]\label{Soundness}
The logic $\mu\text{-}\mathsf{wK4}$ is sound for the class of derivative spaces, and so in particular for the class of weakly transitive frames.
If $\Lambda$ is any extension of $\mathsf{wK4}$, then $\mu\text{-}\Lambda$ is sound for the class of $\Lambda$-spaces, i.e.~the class of spaces validating all theorems of $\Lambda$.
\end{prop}

 \begin{toappendix}

\proof The Necessitation Rule and the Distribution Axiom are sound because of the normality conditions imposed on the derivative operator $d$, while Weak Transitivity is sound due to the weak idempotence of $d$. The soundness of the Fixed Point Axiom and of the Induction Rule follows in the usual way from our (standard) semantics for fixed-point formulas.
The same argument applies to any extension of $\mathsf{wK4}$ and its class of derivative spaces.
\endproof

\end{toappendix}

Our goal is to show that this system is also (weakly) complete, and that the logic is decidable. But for this, we need  first look at some theorems of the above axiomatic system.

\begin{proposition}\label{Theorems}
The following schemas are provable in the logic $\mu\text{-}\mathsf{wK4}$
(for all formulas $\f, \psi$, and formulas $\theta=\theta(x)$ that are positive in $x$):

\begin{enumerate}
\item $\nu x. \theta \, \Leftrightarrow \, \theta(\nu x. \theta)$
\item $(\refbox\f \wedge \theta(\psi))\,  \Rightarrow \, \theta(\refbox\f \wedge \psi)$
\item $\refbox(\f\Rightarrow \psi) \,  \Rightarrow \, (\theta(\f)\Rightarrow \theta(\psi))$
\item $\refbox(\f \Rightarrow \theta(\f)) \,  \Rightarrow \, (\f\Rightarrow \nu x. \theta)$
\end{enumerate}
\end{proposition}

\begin{toappendix}

\begin{proof}%[Proof of Proposition \ref{Theorems}]
Claim 1) is an well-known, easy consequence of the Fixed Point Axiom and the Induction Rule.

For claim 2), it is useful to check first the following special cases:

\begin{description}
\item[2a] $(\refbox\f \wedge \Diamond\psi) \, \Rightarrow \, \Diamond (\refbox\f \wedge\psi)$
\item[2b] $(\refbox\f \wedge \Box\psi) \, \Rightarrow \, \Box (\refbox\f \wedge\psi)$
\item[2c]   $(\refbox\f \wedge \nu x.\theta) \, \Rightarrow \, \nu x.(\refbox\f \wedge \theta)$
\item[2d] $(\refbox\f \wedge \mu x.\theta) \, \Rightarrow \, \mu x.(\refbox\f \wedge \theta)$
\end{description}

Checking that these special instances of 2) follow from the axioms is an easy verification. Given them, one can prove 2) by induction on the complexity of $\theta(x)$, written in positive form (i.e. with negations only in front of propositional variables, other than $x$, and using in rest only conjunctions, disjunctions, $\Diamond$, $\Box$ and the fixed-point operators $\nu x$ and $\mu x$). The atomic cases are immediate, and the inductive steps for conjunction and disjunction follow trivially by propositional logic, while the other inductive steps are taken care by the instance 2)a-2)d above.

To prove claim 3), first note that $(\refbox(\f\Rightarrow\psi)\wedge \f)\Rightarrow \psi$ is a theorem in our axiom system. By using the monotonicity of the positive formula $\theta(x)$ (itself provable in the system), we can derive the theorem $\theta(\refbox(\f\Rightarrow\psi)\wedge \f))\Rightarrow \theta(\psi)$. Putting this together with $(\refbox(\f\Rightarrow\psi)\wedge \theta(\f))\Rightarrow \theta(\refbox(\f\Rightarrow\psi)\wedge \f))$ (which is just a special instance of claim 2)), we obtain $(\refbox(\f\Rightarrow\psi)\wedge \theta(\f))\Rightarrow \theta(\psi)$, from which the desired conclusion follows by propositional reasoning.

Finally, to prove claim 4), we start with the obvious theorem $\refbox(\f\Rightarrow\theta(\f))\Rightarrow (\f\Rightarrow \theta(\f)$, from which we get $(\refbox(\f\Rightarrow\theta(\f))\wedge \f)\Rightarrow \theta(\f)$, and thus also $(\refbox(\f\Rightarrow\theta(\f))\wedge \f) \Rightarrow ( \refbox(\f\Rightarrow\theta(\f))\wedge\theta(\f))$. Putting this together with
$\refbox(\f\Rightarrow\theta(\f))\wedge\theta(\f)) \Rightarrow \theta (\refbox(\f\Rightarrow \theta(\f))\wedge \f)$ (itself an instance of claim 2)), we obtain $(\refbox(\f\Rightarrow\theta(\f))\wedge \f) \Rightarrow
\theta (\refbox(\f\Rightarrow \theta(\f))\wedge \f)$. Applying then the Induction Rule, we derive
$(\refbox(\f\Rightarrow\theta(\f))\wedge \f) \Rightarrow \nu x.\theta$, from which the desired conclusion follows by propositional reasoning.
\end{proof}

 \end{toappendix}

We are now ready to state the first of our main results.
Below, recall that $\Lambda$ has the {\em finite model property} if for any formula $\varphi$, $\varphi$ is a theorem of $\Lambda$ iff $\varphi$ is valid over the class of finite $\Lambda$-models.
The logic $\Lambda$ has the {\em strong finite model property} if the size of a finite countermodel for $\varphi$ can be bounded by a function computable from the length of $\varphi$.

\begin{thm}[Completeness, FMP and Decidability]\label{Completeness}
Let $\mathcal L$ be a natural sublanguage of ${\mathcal L}_\mu$.
The logic $\mu\text{-}\mathsf{wK4}^\mathcal L$ is \emph{(weakly) complete} for the class of all
weakly transitive frames, as well as for the class of Alexandroff spaces (irreflexive weakly transitive frames). Hence, $\mu\text{-}\mathsf{wK4}^\mathcal L$ is complete for the class of all topological spaces, and thus also for the larger class of all derivative spaces. The logic has the \emph{strong finite model property} (wrt all the above classes), and so its satisfiability problem is \emph{decidable}.
\end{thm}

This will be proven in Section \ref{secKripkeProof}, while in Section
\ref{secSubframe} we generalize this result to many other classes of relational structures and the corresponding logics.

We conclude this section by discussing two extensions of $\mathsf{wK4}$ that are of interest in the context of topological semantics.
Recall that a topological space $(\mathcal X,c)$ is $T_0$ if given $x,y\in \mathcal X$ with $x\neq y$, either $x\not\in c\{y\}$ or $y\not\in c\{x\}$ (i.e., the two do not have the same set of neighborhoods).
It is known (see \cite{BEG11}) that the derivational modal logic of $T_0$ spaces is the system
\[\mathsf{wK4T_0} :=
\mathsf{wK4} + p\wedge \Diamond (q\wedge \Diamond p)\to \Diamond p\vee \Diamond (q\wedge \Diamond q).\]
Moreover, $\mathsf{wK4T_0}$ corresponds to the class of $\mathsf{wK4}$ frames $(W,\rel)$ so that $w\rel v\rel w$ implies that $w\rel w$ or $v\rel v$.
Frames satisfying this property are {\em weakly reflexive frames.}
If we define the {\em cluster} of $w\in W$ to be the set of points $v$ so that $v \rrel w \rrel v$
(equivalently: s.t. either $w\simrel v$ or $w=v$), then a weakly transitive frame $(W,\rel)$ is weakly reflexive iff every cluster has at most one irreflexive point.

The second extension we are interested in is $\mathsf{K4}$, given by $\mathsf{wK4}+\ps \ps p\to \ps p$.
It is well known (see, e.g., \cite{CZ97, Esakia04, BEG11}) that this is the logic of all transitive frames, and that it is
also the logic of all {\em $T_D$ spaces}. These are topological spaces $(\mathcal X,c)$ such that every point is isolated in its own closure; i.e., if $x\in \mathcal X$, there is an open set $U$
with $\{x\} = U\cap c\{x\}$.
These results readily extend to the derivative $\mu$-calculus.

\begin{thm}\label{T0,TD}
\begin{enumerate}
\item The logic $\mu\text{-}\mathsf{wK4T_0}$, obtained by adding to $\mathsf{wK4T_0}$ the above Fixed Point Axiom and Induction Rule, is sound and (weakly) complete for the class of all $T_0$ topological spaces.

\item  The logic $\mu\text{-}\mathsf{K4}$, obtained by adding to $\mathsf{K4}$ the above Fixed Point Axiom and Induction Rule, is sound and (weakly) complete for the class of all $T_D$ topological spaces.
\end{enumerate}
\end{thm}

We will prove this result in Section \ref{secTopComp}.
A related completeness result for the $T_D$ case has already been proven in \cite{GH18}.
But that result concerns only the (semantically equivalent) tangled modal logic, while ours is about the full language of $\mu$-calculus.

\medskip

Before proving Theorem \ref{Completeness}, we will make a detour to discuss the tangled derivative in the context of $\sf wK4T_0$ models.
The results in the following section are not needed to establish our main theorems, but they motivate our choice to work over the full $\mu$-calculus rather than focusing on tangled fragments.

\section{Expressive incompleteness of tangle logic}\label{secTan}

A natural question is whether topological $\mu$-calculus collapses to a simpler modal fragment; if so, then a complete axiomatization of the simpler fragment would in principle suffice, and might be easier to prove than for the full calculus. As mentioned in the Introduction, this is exactly what happened on $T_D$ spaces. Dawar and Otto \cite{DO09} showed that the full $\mu$-calculus is expressively equivalent to the so-called \emph{tangled derivative logic }$\mathcal L_{\tangle}$ over the class of (finite) $\sf K4$ frames, and thus also over $T_D$ spaces; while Goldblatt and Hodkinson \cite{GH18} completely axiomatized $\mathcal L_{\tangle}$ over these classes.\footnote{On the other hand, Goldblatt and Hodkinson \cite{GH18} showed that $\tangle$ is not definable in $\mathcal L_{\reftangle}$ over the class of $\sf K4$ frames, and hence over the class of $T_D$ spaces.
It follows that $\mathcal L_{\reftangle}$ is not expressively complete, even over the class of $\sf K4$ frames.}

In this section, we show that the Dawar-Otto result does \emph{not} hold for general spaces, and in fact \emph{not even for $T_0$ spaces}: the tangled derivative logic is no longer expressively equivalent to the $\mu$-calculus over the class of $\sf wK4T_0$ frames, and hence over the class of all $T_0$ spaces.

\medskip

For each finite set of formulas $\Gamma$, consider the \emph{tangled derivative} $\tangle \Gamma$ and \emph{tangled closure} $\reftangle \Gamma$ of $\Gamma$, as defined in Section \ref{secMu}.
Let $\mathcal L_{\tangle}$ and $\mathcal L_{\reftangle}$ be the natural sublanguages of the $\mu$-calculus whose only fixed points are of the respective forms above. To prove that $\mathcal L_{\tangle }$ is not expressively complete for $\mu$-calculus over  $\sf wK4T_0$ frames, we will show that $\reftangle$ is not definable in $\mathcal L_{\tangle}$.

\smallskip

For this, we define a `spine' model $\mathcal S$ based on the ordinal $\omega+3$.
We briefly recall that $\omega$ denotes the first infinite ordinal, and follow the set-theoretic convention that each ordinal is identified with its set of predecessors.
We moreover use interval notation on the ordinals: $(\alpha,\beta)$ is the set of ordinals $\xi$ with $\alpha<\xi<\beta$.

With this in mind, we set $\mathcal S = (\omega+3,\rel,\|\placeholder\|)$, where
\begin{enumerate}

\item $\alpha \rel \beta$ if one of the following occurs:
\begin{enumerate}

\item $\alpha > \beta$;

\item $\alpha=\beta$ and $\alpha$ is odd (including $\omega+1$), or

\item $\alpha = \omega+1$ and $\beta=\omega+2$.

\end{enumerate}

\item $\alpha \in \|p\|$ iff $\alpha$ is odd, $ \|q\| = \varnothing$ for all $q\neq p$.

\end{enumerate}

\begin{lemma}\label{lemmIsModel}
$\mathcal S$ is a $\sf wK4T_0$ model.
\end{lemma}

 \begin{toappendix}

\begin{proof}%[Proof of Lemma \ref{lemmIsModel}]
Weak transitivity is easily checked using a case distinction.
The $T_0$ condition is checked by noting that all clusters are singletons, except for $\{\omega+1,\omega+2\}$.
But only $\omega+2$ is irreflexive, as needed.
\end{proof}

 \end{toappendix}

The goal is to show that over $\mathcal S$, no $\mathcal L_{\tangle}$ formula is equivalent to $\reftangle\{p,\neg p\} $.
First, we evaluate the latter.
%, and $\rel $ for the strict part of $ \crel$.

\begin{lemma}\label{lemmEvalTang}
Over $\mathcal S$, $\| \reftangle\{p,\neg p\} \| = \{\omega+1,\omega +2\}$.
\end{lemma}

 \begin{toappendix}

\begin{proof}%[Proof of Lemma \ref{lemmEvalTang}]
We have that $\alpha \in \| \reftangle\{p,\neg p\} \|$ if and only if there is an infinite chain
\[\alpha  \crel  \beta_0 \crel  \beta_1 \crel  \beta_2 \crel \ldots\]
such that $\beta_i \not \in V(p) $ when $i$ is even, $\beta_i  \in V(p) $ when $i$ is odd.
From the latter it follows that $\beta_i \neq \beta_{i+1}$.
Since $\omega+3$ with the usual ordering is well-founded, such a chain can only occur in the ``ill-founded'' part of our model, namely $\{\omega+1,\omega+2\}$.
However, the infinite chain
\[\omega+2 \crel \omega+1 \crel \omega+2 \crel \ldots \]
witnesses that $\{\omega+1,\omega+2\} \subseteq \|  \reftangle\{p,\neg p\} \|$.
\end{proof}

 \end{toappendix}

\begin{lemma}\label{lemmEvenOdd}
If $\varphi$ is any formula of $\mathcal L_{\tangle}$ then there exists $n_\varphi  < \omega$ such that for every $\alpha,\beta \in (n_\varphi,\omega+3)$ which are either both even or both odd, $\alpha \in \|\varphi\|$ iff $\beta \in \|\varphi\|$.
\end{lemma}

 \begin{toappendix}

\begin{proof}
By induction on the complexity of $\varphi$.
The base case follows from the definition of $\|\placeholder\|$ and the cases for Booleans are straightforward.
Consider the case where $\varphi = \Diamond\psi$.
By the induction hypothesis, $n_\psi$ is well-defined and finite, and we can take $n_\varphi = n_\psi+2$.
Then, if $\alpha,\beta > n_\varphi$ and $\alpha\in \|\Diamond\psi\|$, there is $\alpha'$ such that $\alpha \rel  \alpha'$ and $\alpha' \in \| \psi\|$.
If $\alpha'<n_\psi$, set $\beta'=\alpha'$; otherwise, take $\beta'\in \{n_\psi + 1, n_\psi +2\}$ of the same parity as $\alpha'$.
We then see that $\beta \rel  \beta'$, so that $\beta \in \|\Diamond\psi\|$.

Finally, consider $\varphi = \tangle \Gamma$.
We may assume that $\Gamma\neq\varnothing$, since $\tangle \varnothing$ is tautologically true.
Let $n_\varphi = \max_{\gamma\in \Gamma} n_\gamma+2$.
Suppose that $\alpha,\beta > n_\varphi$, and that $\alpha\in \|\tangle \Gamma \|$.
Let $\alpha_*$ be the least element of $\|\tangle \Gamma \|$ with respect to the usual ordinal ordering.
First assume that $\alpha_*<\omega+1$.
Then, for all $\gamma\in \Gamma$, there is $\xi_\gamma \in \|\gamma \wedge \tangle \Gamma\|$ such that $\alpha_*\rel  \xi_\gamma $.
But then, by the definition of $\rel$ and the minimality of $\alpha_*$ we must have that $\xi_\gamma = \alpha_*$, and thus $\alpha_*$ satisfies every element of $\Gamma$.
Note also that $\alpha_*$ is reflexive, so $\alpha_*$ must be odd.

It follows that $\alpha_* \leq n_\varphi  $, since otherwise the odd element of $\{ n_\varphi-1, n_\varphi \}$ satisfies all formulas of $\Gamma$ and hence $\tangle \Gamma$, contradicting the minimality of $\alpha_*$.
But then, from $\beta>n _\varphi$ we see that $\beta \rel  \alpha_* $, so $\beta\in \| \tangle \Gamma \|$.

Finally, we consider the case where $\alpha_* \geq \omega +1$.
In fact, we will show that this case is impossible.
Note that in this case $\omega+2 \in \| \tangle \Gamma\|$.
As before, for all $\gamma\in \Gamma$, there is $\xi_\gamma \in \|\gamma \wedge \tangle \Gamma\|$ such that $\omega + 2 \rel  \xi_\gamma $.
But by minimality of $\alpha_*$, the only option is to have $\xi_\gamma = \omega+1$, so in fact $\omega+1$ satisfies all elements of $\Gamma$.
Reasoning as above, letting $\alpha'\in \{n_\varphi-1, n_\varphi \}$ be odd, we see that $\alpha'$ satisfies all formulas of $\Gamma$.
But then $\alpha' \in \|\tangle\Gamma\|$, contradicting the choice of $\alpha_*$.
\end{proof}

\begin{remark}
From the proof it can be estimated that it suffices to take $n_\varphi = 2|\varphi|$.
\end{remark}

 \end{toappendix}

Since $ n<\omega,\omega+2$ for all $n<\omega$ and the two are even, we obtain the following special case.

\begin{corollary}
In $\mathcal S$, $\omega$ and $\omega+2 $ satisfy the same formulas of $\mathcal L_{\tangle}$.
\end{corollary}

However, we have seen that $\omega+2 \in \|\reftangle \{p,\neg p\}\|$, but $\omega \not \in \|\reftangle \{p,\neg p\}\|$.
We may thus conclude that $\reftangle \{p,\neg p\}$ is not definable.

\begin{thm}
The formula $\varphi = \reftangle \{p,\neg p\}$ is not definable in $\mathcal L_{\tangle}$, even by an infinite set of formulas.
\end{thm}

Given that $ \reftangle\{p,\neg p\}$ is definable in the $\mu$-calculus but not in $\mathcal L_{\tangle}$, we obtain the following result.

\begin{corollary}
Not every formula of the $\mu$-calculus is definable in $\mathcal L_{\tangle}$ over the class of $\sf wK4T_0$ frames.
\end{corollary}

For this reason, in this paper we will work over the full $\mu$-calculus, rather than the tangled derivative fragment.

\section{Truth-preserving maps}\label{secMorphism}

In this section, we focus on the relational semantics, and review and generalize some well-known properties of $\mu$-calculus \cite{Venema08}: locality and invariance under bounded morphisms.

\medskip

\begin{definition}[D-morphisms and $P$-morphisms]
A {\em d-morphism} between derivative spaces $(\mathcal{X}, d)$ and $(\mathcal{X'},d')$ is a function $\pi\colon \mathcal{X}\to \mathcal{X'}$ such that $\pi^{-1}d'(X')= d\pi^{-1}(X')$ for all sets $X'\subseteq \mathcal{X'}$.

If $\pi$ is surjective, we say that the space $\mathcal{X'}$ is a \emph{d-morphic image} of the space $\mathcal{X}$.

For any set $P\subseteq \pvar$, a \emph{$P$-morphism} between derivative models $\bM=(\mathcal{X},d,\|\placeholder\|)$ and $\bM'=(\mathcal{X'},d',\|\placeholder\|')$ is a d-morphism $\pi\colon \mathcal{X}\to \mathcal{X'}$ s.t.
$\pi^{-1}\|x\|'=\|x\|$ for all atoms $x\in P$. 

If $\pi$ is surjective, we say that the model $\bM'$ is a \emph{$P$-morphic image} of the model $\bM$.
\end{definition}

\begin{remark}
The notion of $P$-morphism is a generalization to derivative spaces of the well-known concept of {\em p-morphism} \cite{BdRV01}, albeit relativized to a set of variables $P$.
The restriction to such a set of variables (particularly, when $P$ is finite) will be essential in many of our proofs.
\end{remark}

\begin{lemma}\label{lemmaDmorVal}
If $\pi\colon \mathcal{X}\to \mathcal{X'}$ is a $P$-morphism between derivative models $\bM=(\mathcal{X},d,\|\placeholder\|)$ and $\bM'=(\mathcal{X'},d',\|\placeholder\|')$, then for all $\mu$-calculus formulas $\f=\f(\tuple{x})\in\mathcal{L}_\mu^P$ and tuples of sets $\tuple{X}, \tuple{X'}$ s.t. $X_i=\pi^{-1}X'_i$ for all $i$, we have:
$$\|\f(\tuple{X})\|=\pi^{-1}\|\f(\tuple{X'})\|'.$$
\end{lemma}

 \begin{toappendix}

\proof%[Proof of Lemma \ref{lemmaDmorVal}]
Induction on the complexity of the formula $\f=\f(\tuple{x})$. The atomic case $\f=x_i$ follows immediately from the assumption that $X_i=\pi^{-1}X'_i$, while the atomic case $\f=y$ for $y\in P$ not occurring in $\tuple{x}$ follows from atomic requirement on $P$-morphisms. The Boolean cases follow from two well-known properties of the inverse map: $\pi^{-1}(\mathcal{X'}-X')=\mathcal{X}-\pi^{-1}X'$ and $\pi^{-1}(X'\cap Y')=\pi^{-1}X'\cap \pi^{-1}Y'$. The derivative case $\Diamond\f$ is an immediate consequence of the definition of $d$-morphism.

\emph{The case $\nu y. \f(y,\tuple{x})$}. We assume the induction hypothesis for $\f$, and we first prove the following

\emph{Claim}: If $\f^\alpha_y(\tuple{X})$ is the transfinite sequence of sets in Proposition \ref{Mono}(3), then for all ordinals $\alpha$ we have
$$\pi^{-1}\f^\alpha_y(\tuple{X'})=\f^\alpha_y(\tuple{X}).$$

We prove this Claim by subinduction on $\alpha$:\\ $\pi^{-1}\f^\alpha_y(\tuple{X'})=\pi^{-1}(\bigcap_{\beta<\alpha} \|\f(\f^\beta_y(\tuple{X'}),\tuple{X'})\|')=
\bigcap_{\beta<\alpha} \pi^{-1} \|\f(\f^\beta_y(\tuple{X'}),\tuple{X'})\|'=
\bigcap_{\beta<\alpha} \|\f(\f^\beta_y(\tuple{X}),\tuple{X})\|=
\f^\alpha_y(\tuple{X})$ (where at the third step we used both the induction hypothesis for $\f$ and the subinduction hypothesis for $\beta$, as well as the background assumption about $\tuple{X},\tuple{X'})$.

Given the Claim, we can now prove the inductive case for $\nu y. \f(y,\tuple{x})$:\\
$\pi^{-1}\|\nu y.\f(y,\tuple{X'})\|'= \pi^{-1} (\bigcap_\alpha \f^\alpha_y(\tuple{X'}))=
\bigcap_\alpha \pi^{-1} \f^\alpha_y(\tuple{X})= \|\nu y.\f(y,\tuple{X})\|$. \endproof

 \end{toappendix}

It is useful to keep in mind the special case where the tuple of substitution variables is empty.

\begin{cor}[Invariance under $P$-morphisms]\label{cor1DmorVal}
If $\pi\colon \mathcal{X}\to \mathcal{X'}$ is a $P$-morphism between derivative models $\bM=(\mathcal{X},d,\|\placeholder\|)$ and $\bM'=(\mathcal{X'},d',\|\placeholder\|')$, then for all  $\f\in\mathcal{L}_\mu^P$ we have:
$$\|\f\|=\pi^{-1}\|\f\|'.$$
\end{cor}

 \begin{toappendix}
\proof%[Proof of Corollary \ref{cor1DmorVal}]
Apply Lemma \ref{lemmaDmorVal} to the empty tuple of variables, and corresponding empty tuples of sets. (Alternatively: let $\tuple{x}=(x_1, \ldots, x_n)$ be a tuple enumerating all the free variables in $\f$. Then apply Lemma \ref{lemmaDmorVal} to this tuple, and to tuples of sets $\tuple{X}:=(\|x_1\|, \ldots, \|x_n\|)$ and
$\tuple{X'}:=(\|x_1\|', \ldots, \|x_n\|')$.)\endproof
 \end{toappendix}

\medskip

In practice, $P$-morphisms are most useful when they are surjective, as they then preserve validity of formulas.

\medskip

\begin{cor}\label{cor2DmorVal}\
\begin{enumerate}
\item If a derivative model $\bM'=(\mathcal{X'},d',\|\placeholder\|')$ is a $P$-morphic image of a model $\bM=(\mathcal{X},d,\|\placeholder\|)$, then the two models validate (satisfy) the same formulas of $\mathcal{L}_\mu^P$.
\item If a derivative space $(\mathcal{X'},d')$ is a d-morphic image of space $(\mathcal{X},d)$, then every formula that is satisfiable on $(\mathcal{X'},d')$ is also satisfiable on $(\mathcal{X},d)$; equivalently: every formula that is valid on $(\mathcal{X},d)$ is also valid on $(\mathcal{X'},d')$.
\end{enumerate}
\end{cor}

 \begin{toappendix}

\proof%[Proof of Corollary \ref{cor2DmorVal}]
To check part 1, we prove the satisfiability version. Let $\f\in \mathcal{L}_\mu^P$ and let $\pi:\mathcal{X}\to \mathcal{X'}$ be a surjective $P$-morphism.  By Corollary \ref{cor1DmorVal}, we have $\|\f\|=\pi^{-1}\|\f\|'$. Combining this with the functionality and surjectivity of $\pi$, we obtain the equivalence: $\|\f\|\not=\varnothing$ iff $\|\f\|'\not=\varnothing$.

For part 2: we again check the satisfiability version. Let $\pi:\mathcal{X}\to \mathcal{X'}$ be a surjective d-morphism, and let $\f$ be satisfiable on $(\mathcal{X'},d')$, i.e. there exists some valuation $\|\placeholder\|'$ satisfying $\f$ at some point of $\mathcal{X'}$.
Take the map $\|\placeholder\|:=\pi^{-1}\|\placeholder\|'$ defined on $\mathcal{X}$. Then  $\|\placeholder\|$ is a valuation on $\mathcal{X}$ that makes $\pi$ a surjective $\pvar$-morphism, hence by part 1, $\|\placeholder\|$ satisfies $\f$ at some point of $\mathcal{X}$.\endproof

 \end{toappendix}

\medskip

It is useful to have a more `bisimulation-like' characterization of d-morphisms.
Using the equivalence between derivative spaces and neighborhood derivative spaces, we can characterize d-morphisms in terms of d-neighborhoods:

\begin{lemma}\label{lemmDmorDeriv}
Let $\pi:\mathcal{X}\to \mathcal{X'}$ be a map between derivative spaces $(\mathcal{X},d)$ and $(\mathcal{X'},d')$. Then the following are equivalent:

\begin{enumerate}
\item $\pi$ is a d-morphism;
\item the conjunction of the following back-and-forth conditions holds for all points $x\in \mathcal{X}$ and all sets $X\subseteq \mathcal{X}$ and $X'\subseteq \mathcal{X'}$:
\begin{itemize}
\item (\emph{back}) $X' \in \mathcal{N}_{d'}(\pi(x))$ implies $\pi^{-1}(X')\in  \mathcal{N}_d(x)$, and
\item (\emph{forth}) $X\in \mathcal{N}_d(x)$ implies $\pi(X)\in \mathcal{N}_{d'}(\pi(x))$;
\end{itemize}
\item $\pi^{-1}(X')\in \mathcal{N}_d(x)$ iff $X'\in \mathcal{N}_{d'}(\pi(x))$, for all $x\in \mathcal{X}$ and $X'\subseteq \mathcal{X'}$.
\end{enumerate}
\end{lemma}

This follows from the general theory of bounded morphisms in monotonic neighborhood models \cite{Hans03,Pac17}: indeed, the third equivalent statement is exactly the definition of a bounded morphism in monotonic neighborhood semantics. When both spaces are topological derivative spaces, the back-and-forth conditions refer to punctured neighborhoods. When both are topological closure spaces, we obtain the usual notion of \emph{interior map}. The case where $\mathcal{X}$ is a topological space and $\mathcal{X'}$ a Kripke frame is of a special interest:

\begin{cor}\label{lemmDmorKripke}
Let $\pi:\mathcal{X}\to \mathcal{X'}$ be a map between a topological derivative space
$(\mathcal{X},d)$ and a weakly transitive frame $(\mathcal{X'},\rel)$. Then the following are equivalent:

\begin{enumerate}
\item $\pi$ is a d-morphism;
\item the conjunction of the following back-and-forth conditions holds for all points $x\in \mathcal{X}$:
\begin{itemize}
\item $\pi(U-\{x\})\subseteq \pi(x){\uparrow}$, for \emph{some} open neighborhood $U$ of $x$, and

\item $\pi(x){\uparrow} \subseteq \pi(U- \{x\})$, for \emph{all} open neighborhoods $U$ of $x$.
\end{itemize}
\end{enumerate}
\end{cor}

Finally, when both spaces are weakly transitive frames, we recover the standard notion of bounded frame morphism:

\smallskip

\par\noindent\textbf{$P$-morphisms and $P$-bisimulations between relational models.} When both $\mathcal{X}$ and $\mathcal{X'}$ are weakly transitive frames, it is easy to see that our notion of $P$-morphism matches the standard modal notion of \emph{p-morphisms} (also known as ``bounded $P$-morphisms"), i.e. \emph{functional $P$-bisimulations}.

\begin{definition}
Let $\bM_1=(W_1,\rel_1,\|\cdot\|_1)$ and $\bM_2=(W_2,\rel_2,\|\cdot\|_2)$ be relational models.
A relation $B\subseteq W_1\times W_2$ is a \emph{$P$-bisimulation} if, for all states $w_1\in W_1, w_2\in W_2$, $(w_1,w_2)\in B$ implies three conditions: (a) $w_1\in\|p\|_1$ iff $w_2\in \|p\|_2$ (\emph{Atomic Preservation}); (b) if $w_1\rel_1 s_1$ then there exists some $s_2\in W_2$ with $w_2\rel_2 s_2$ and $(s_1,s_2)\in B$ (\emph{Forth} Condition); (c) if $w_2 \rel_2 s_2$ then there exists some $s_1\in W_1$ with $w_1 \rel_1 s_1$ and $(s_1,s_2)\in B$ (\emph{Back} Condition).
\end{definition}

Then, a bounded $P$-morphism is just a functional $P$-bisimulation.
It is well known that relational $P$-bisimulations between weakly transitive relational models $\bM_1=(\mathcal{X}_1,\rel_1, \|\placeholder\|_1)$ and $\bM_2=(\mathcal{X}_2,\rel_2, \|\placeholder\|_2)$ are exactly the relations of the form $\pi_1^{-1};\pi_2$, where $\pi_1:\bM\to\bM_1$ and $\pi_2:\bM\to \bM_2$ are $P$-morphisms from some other weakly transitive model $\bM$ into the two models, and $;$ is relational composition.\footnote{This relationship between $P$-bisimulations and spans of bounded $P$-morphisms is well-known in modal logic, and has lead to the general definition of coalgebraic bisimulation, as a span of coalgebraic morphisms.}
\medskip

\par\noindent\textbf{Invariance under bisimilarity} The relation of \emph{$P$-bisimilarity} $\simeq_P$ on a given model $\bM=(W, \rel,  \|\placeholder\|)$ is the largest $P$-bisimulation relation ${\simeq_P} \subseteq W\times W$. When $P=\pvar$, we drop the subscript, writing e.g. $s\simeq w$ and talking simply of `bisimulation' and `bisimilarity'. It is easy to see that $P$-bisimilarity is an equivalence relation on $W$. The following fact is a widely known feature of $\mu$-calculus:

\begin{prop}[Invariance under Bisimilarity]\label{Bisim}
The valuation $\|\f\|$ of every formula $\f \in \mathcal{L}_\mu^P$ is closed under $P$-bisimilarity: for all
$s,w\in W$, if $s\simeq_P w$ and $s\in \|\f\|$, then $w\in\|\f\|$.
\end{prop}

%\begin{toappendix}
\proof
%[Proof of Proposition \ref{Bisim}]
This is well-known (and easy to verify directly).
\endproof
%\end{toappendix}

\medskip

\par\noindent\textbf{Locality}
Another known fact is that $\mu$-calculus is ``local": the truth value of a formula $\f$ at a state depends only on the accessible part of the model (i.e., the so-called generated submodel). This can be generalized as follows:

\begin{lemma}\label{locality}
Let $\f=\f(\tuple{x},\tuple{y})$ be a formula. Then we have the following:
\begin{enumerate}
\item $\|\f(\tuple{X},\tuple{Y})\|\cap w\rupst= \|\f(\tuple{X},\tuple{Y\cap w\rupst})\|\cap w\rupst$, for all states $w\in W$ and tuples of sets of states $\tuple{X},\tuple{Y}$;
\item If $\tuple{y}=(\tuple{z}, y)$ and $\f=\f(\tuple{x},\tuple{y})=\f(\tuple{x},\tuple{z}, y)$ is positive in $y$, then
    $$\f_y^\alpha(\tuple{X},\tuple{Z})\cap w\rupst=\f_y^\alpha(\tuple{X},\tuple{Z\cap w\rupst})\cap w\rupst$$
    for all states $w\in W$, ordinals $\alpha\in {\sf On}$ and tuples of sets of states $\tuple{X},\tuple{Z}$.
(Here, $\f_y^\alpha$ is the sequence introduced in Proposition \ref{Mono}(3).)
\end{enumerate}
\end{lemma}

 \begin{toappendix}
\proof%[Proof of Lemma \ref{locality}]
We show the two claims by simultaneous\david{Added `simultaneous', I think this is what was meant.} induction on the subformula-complexity of $\f$. For claim (1), the base case $\f=x$, as well as the inductive case for Boolean operators, are trivial.

\emph{The case of $\Diamond \f$ for (1)}: $\|\Diamond\f (\tuple{X},\tuple{Y})\|\cap w\rupst=
\{w\in W : \exists s\in w{{\uparrow}}  \mbox{ s.t. } s\in \|\f(\overline{X}, \tuple{Y})\|\}\cap  w\rupst$. By the induction hypothesis (for $\f$ and $s\in w{{\uparrow}}\subseteq w\rupst$), this is equal to $\{w\in W : \exists s\in w{{\uparrow}} \mbox{ s.t. } s\in \|\f(\overline{X}, \tuple{Y\cap w\rupst})\|\}\cap  w\rupst$, i.e. to $\|\Diamond\f(\tuple{X},\tuple{Y\cap w\rupst})\|\cap w\rupst$, as desired.

\emph{The case of $\nu z. \f$ for (1)}: Using Proposition \ref{Mono}(3) and the inductive hypothesis (2) for $\f$, we have:
$\|\nu z.\f (\tuple{X},\tuple{Y})\|\cap w\rupst =
\bigcap_{\alpha\in On} \f_z^\alpha (\tuple{X},\tuple{Y}) \cap  w\rupst =
\bigcap_{\alpha\in On} \f_z^\alpha (\tuple{X},\tuple{Y\cap w\rupst}) \cap  w\rupst =
\|\nu z.\f (\tuple{X},\tuple{Y\cap w\rupst})\|\cap w\rupst$.

\emph{To prove claim (2) for $\f$}, assume claim (1) for $\f=\f(\tuple{x},\tuple{z}, y)$ (for all set tuples), and prove (2) by subinduction on the ordinal $\alpha$:
\begin{align*}
\f_y^\alpha & (\tuple{X},\tuple{Z})\cap w\rupst  =
\bigcap_{\beta< \alpha} \|\f(\tuple{X},\tuple{Z}, \f_y^\beta(\tuple{X},\tuple{Z}))\|
\cap w\rupst \\
& =
\bigcap_{\beta< \alpha} \|\f(\tuple{X},\tuple{Z\cap w\rupst}, \f_y^\beta(\tuple{X},\tuple{Z}))\|
\cap w\rupst\\
& =
\bigcap_{\beta< \alpha} \|\f(\tuple{X},\tuple{Z\cap w\rupst}, \f_y^\beta(\tuple{X},\tuple{Z\cap w\rupst}))\|
\cap w\rupst\\
& =
\f_y^\alpha(\tuple{X},\tuple{Z\cap w\rupst})\cap w\rupst,
\end{align*}
where we used first the induction hypothesis for $\f$, then the subinduction hypothesis for $\beta<\alpha$.
\endproof
 \end{toappendix}

\medskip

\par\noindent\textbf{Asserting properties locally above a point} Given a point $w\in W$, and given a property $P(X_1, \ldots, X_n)$ involving sets $X_1, \ldots, X_n\subseteq W$, we say that \emph{$P(X_1, \ldots, X_n)$ holds above $w$} if we have $P(X_1\cap w\rupst, \ldots, X_n\cap w\rupst)$. In particular, for two sets $X,Y\subseteq W$, we say that \emph{$X=Y$ holds above $w$} iff $X\cap w\rupst=Y\cap w\rupst$.

\medskip

\par\noindent\textbf{Depth of a point in a model} Recall that $\srel$ is the strict preorder induced by $\rel$. Given a weakly transitive model $\bM=(W, \rel , \|\placeholder\|)$, and a point $w\in W$, a \emph{strict (finite) $w$-chain} is a finite sequence of points of the form $w=w_0 \srel w_1 \srel \ldots \srel w_n$. The number $n$ is called the \emph{length} of our finite chain. The \emph{depth $\dpt (w)$ of the point $w\in W$} is the supremum of the lengths of all strict $w$-chains.
In general, we have $\dpt (w)\geq 0$, with $\dpt (w)=0$ iff for every $s\in W$, $w \rel s$ implies $s \rel w$; and $\dpt (w)=\omega$ iff there exist $w$-chains of every length $n\in \mathbb N$.
The \emph{depth $\dpt (\bM)$ of the model} $\bM$ is the supremum of the depths of all points of the model:
$$\dpt (\bM):= \sup\{\dpt (w): w\in W\}.$$

\begin{lemma}\label{Depth}
Let $\bM=(W, \rel, \|\placeholder\|)$ be a weakly transitive model, and $w, s\in W$ be two points.
Then we have the following:
\begin{enumerate}
\item if $w \rrel s$, then $\dpt (w)\geq \dpt (s)$;
\item if $w \simrel s$, then $\dpt (w)=\dpt (s)$;
\item if $w \rel s$ and $\dpt (w)=\dpt (s)<\omega$, then $w \simrel s$;
\item if $w \srel s$ and $\dpt (s)$ is finite, then $\dpt (w)>\dpt (s)$.
\end{enumerate}
\end{lemma}

%\begin{toappendix}
\proof
%[Proof of Lemma \ref{Depth}]
Easy verification.
\endproof
%\end{toappendix}

\medskip

Our goal in the next section is to prove Proposition \ref{Completeness}, in particular the completeness of our axiomatization with respect to irreflexive, weakly transitive frames. But for this, recall first that modal logic cannot express irreflexivity. The following result allows us to drop the irreflexivity condition:

\begin{lemma}\label{Irref-drop}
For every weakly-transitive model $\bM$, there exists some irreflexive weakly-transitive model $\widetilde{\bM}$ that validates/satisfies the same $\mu$-calculus formulas as $\bM$. Moreover, if $\bM$ is finite, then $\widetilde{\bM}$ can be taken to be finite as well.
\end{lemma}

\begin{toappendix}
\begin{proof}%[Proof of Lemma \ref{Irref-drop}]
Given any weakly transitive model $\bM=(W,\rel , \| \cdot  \|)$, we associate to it an irreflexive and weakly transitive model $\widetilde{\bM}=(\widetilde{W}, \widetilde{\rel}, \widetilde{\|\cdot\|})$, by first taking
\begin{align*}
\widetilde{W}:= &\{x\in W: x\text{ is irreflexive}\} \\
 &\cup \{(x,i)\in W\times \{0,1\}: x\rel x\}.
\end{align*}
It is useful to consider a map $\pi:\widetilde{W}\to W$, given by $\pi(x,i):=x$ (for reflexive points $x\in W$) and $\pi(x):=x$ (for irreflexive points $x\in W$). Using this, we can define the accessibility relation on $\widetilde{W}$ by putting
$$\widetilde{x}  \widetilde{\rel} \widetilde{y} \, \, \, \mbox{ if } \,\, \, \pi(\widetilde{x}) \rel \pi(\widetilde{y}) \mbox{ and } \widetilde{x}\not=\widetilde{y},$$
for all $\widetilde{x}, \widetilde{y}\in \widetilde{W}$; and we define the valuation on $\widetilde{W}$ by
$$\widetilde{\|p\| }:=\{\widetilde{x}\in \widetilde{W}: \pi(\widetilde{x})\in \|p\|\}.$$
It is easy to see that $\widetilde{\bM}$ is an irreflexive and weakly transitive relational model, and that the map $\pi: \widetilde{W} \to W$ is a $\pvar$-morphism. Since by Proposition \ref{Bisim}, all formulas of $\mu$-calculus are invariant under bisimulation, the two models are equivalent with respect to our syntax.
\end{proof}
 \end{toappendix}

So, to prove Proposition \ref{Completeness}, it is enough to show \emph{completeness and FMP for weakly transitive frames}. This is topic of the next section.

\section{Proof of the main Completeness/FMP result}\label{secKripkeProof}

In this section, we prove our main completeness result (Theorem \ref{Completeness}).
Throughout the section, we fix a consistent formula $\f_0$, and let $P_0 = \free(\f_0)$.
We also fix some
\emph{finite} set $\Sigma\subseteq \mathcal{L}_\mu$, with the following properties: $\f_0\in\Sigma$;
$\Sigma$ is closed
under subformulas; $\Sigma$ is closed up to logical equivalence (in our axiomatic system) under negation $\neg\f$ and under $\refdia\varphi$ operators. The existence of such a finite set $\Sigma$ (for every formula $\f_0$) follows from the fact that $\refdia$ is provably an $\sf S4$-type modality, together with the well-known fact that there are only finitely many non-equivalent modalities in the modal system $\sf S4$ \cite[Ch.~3]{CZ97}.
Note that $\f_0\in\Sigma$, and $P_0 \subseteq\Sigma$ is finite.

\bigskip

\par\noindent\textbf{Plan of the Proof} We start with the canonical model $\Omega$ (comprising all maximally consistent theories), a standard construction in modal logic. But we should stress that the \emph{canonical model is not our intended model}. Indeed,  the usual Truth Lemma \emph{fails} for the $\mu$-calculus in the canonical model: consistent $\mu$-calculus formulas are \emph{not} necessarily satisfied in the canonical model by the theories that contain them.\footnote{To see this, consider atoms $(p_n)_{n <\omega}$ and check that for every $n$, the set $\Phi_n:=\{p_n, \neg\Diamond^\infty \top\}\cup\{\refbox (p_i\Rightarrow \Diamond p_{i+1}): i <\omega\}$ is consistent (since all finite subsets are satisfiable). Use the Canonical Truth Lemma for Basic Modal Logic (and the fact that $\refbox$ is definable in it) to construct $(T_n)_{n <\omega}$ with $\Phi_n\subseteq T_n$ and $T_0\to T_1 \to\ldots \to T_n\to\ldots$. Thus, $T_0\models \Diamond^\infty\top$ although $(\neg \Diamond^\infty\top)\in T_0$.} In fact, the notion of truth in the canonical model will play \emph{no role} in this paper: we never evaluate our formulas in $\Omega$. Instead, we only use a few basic syntactic properties of this model.

Next, we select a special submodel of the canonical model $\Omega^\Sigma$ (called the \emph{$\Sigma$-final model}). Essentially, this consists of the theories whose cluster is locally definable by some formula in $\Sigma$. Our goal will be to show that the Truth Lemma \emph{does} hold in $\Omega^\Sigma$ for $\Sigma$-formulas. It is easy to show that $\Omega^\Sigma$ satisfies the usual $\Diamond$-Witness Lemma for formulas in $\Sigma$, but extending this to fixed points requires some work.

An important role will be played by the notion of $\Sigma$-bisimilarity, a strengthening of the standard notion of bisimilarity, in which the Atomic Permanence clause is replaced by the requirement that $\Sigma$-bisimilar theories agree on $\Sigma$-formulas.
Since it is stronger than usual $P_0$-bisimilarity, $\Sigma$-bisimilarity still preserves the truth values of $\mu$-calculus formulas, as long as their free variables belong to $\Sigma$.

Another key ingredient in our proof is the fact that $\Omega^\Sigma$ is ``essentially" a finite object: though possibly infinite in size, it has finite `depth', and moreover it contains only finitely many $\Sigma$-bisimilarity classes. As a consequence, all relevant fixed points are attained at some fixed finite stage of the iterative process from Proposition \ref{Mono}(3).

We will then use these ingredients to prove our Truth Lemma for the final model $\Omega^\Sigma$. The inductive step for the fixed-point formulas uses the fact that the valuation of these formulas is locally definable by some $\Sigma$-formula.

Once completeness is obtained in this way, we will prove the finite model property by taking the quotient of the final model $\Omega^\Sigma$ modulo $\Sigma$-bisimilarity.

\bigskip

\par\noindent\textbf{Canonical Model} The standard `canonical model' construction provides an (infinite) weakly transitive model. A \emph{theory} is a maximally consistent set of formulas in $\mathcal{L}_\mu$ (i.e. a set $T\subseteq \mathcal{L}_\mu$ that is consistent and has no proper consistent extension). We denote by $\Omega$ the family of all
theories. The \emph{canonical accessibility relation} $\access$ between two such theories
$T, T'\in \Omega$ is given as usual by putting
$$T\access T' \,\,\mbox{ iff } \,\, \forall \varphi \, \left(\mbox{ if } \Box\varphi \in T \mbox{ then } \varphi \in T'\right),$$
and the canonical valuation is given by
$$\|x\|\,\, :=\,\, \{T\in \Omega : x\in T\}.$$
The \emph{canonical model} is the structure $(\Omega, \access, \|\placeholder\|)$. Since the weak-transitivity condition is Sahlqvist, it immediately follows that \emph{the canonical model is weakly transitive} (though not irreflexive); see \cite{BdRV01,CZ97} for details on Sahlqvist formulas and their properties. As a consequence, \emph{the reflexive closure,} which we denote $\crel $, of the canonical relation coincides with its reflexive-transitive closure.

We will make use of a few well-known properties of the canonical model, given by the next four results (see, e.g., \cite{BdRV01}).

\begin{lemma}[Lindenbaum Lemma]\label{lindenbaum}
Every consistent set $\Phi$ of formulas can be extended to a maximal consistent set $T\in \Omega$ s.t.~$\Phi\subseteq T$.
\end{lemma}

\begin{lemma}[Canonical $\Diamond$-Witness Lemma]\label{existence}
For every theory $T\in \Omega$ and formula $\varphi\in \mathcal{L}_\mu$, we have that $\Diamond\varphi\in T$ iff there exists some theory $T'\in \Omega$ s.t. $T\access T' \ni\varphi$.

We also have an equivalent statement in $\Box$-form:
$$\Box\varphi\in T \,\, \mbox{ iff } \,\, \forall T'\in \Omega \left(\mbox{ if } T\access T' \mbox{ then } \varphi\in T'\right).$$
\end{lemma}

The left-to-right implication in the first statement above is known as the (Canonical) $\Diamond$-Existence Lemma.
The proofs are well-known (see, e.g., \cite[Ch.~4]{BdRV01}), and these results imply that the so-called Truth Lemma holds in the canonical model for the $\Diamond$-fragment of our logic.

In fact, we can extend this to a Canonical $\refdia$-Witness Lemma, using the following result

\begin{lemma}\label{BoxStar}
For theories $T, T'\in \Omega$, we have:
$$T\crel  T' \,\, \mbox{ iff } \,\, \forall \varphi (\mbox{ if } \refbox\varphi\in T \mbox{ then } \varphi\in T').$$
%where  $\crel = \access\cup id$ is the reflexive closure of $\access$.
\end{lemma}

 \begin{toappendix}
\proof%[Proof of Lemma \ref{BoxStar}]
The \emph{left-to-right implication}: Assume that $T\crel  T'$. If $T=T'$, then $\refbox\varphi\in T$ implies by definition that $\varphi \in T=T'$, as desired. If $T\not=T'$, then we must have $T\access T'$, and then $\refbox\varphi\in T$ implies by definition that $\Box\varphi \in T$, which implies that $\varphi\in T'$ (by the Canonical $\Diamond$-Witness Lemma), as desired.

The \emph{right-to-left implication}: Assume that we have $\forall \varphi (\refbox\varphi\in T\Longrightarrow \varphi\in T')$. To show that $T\crel  T'$, we assume that $T\not=T'$, and we need to prove that $T\access T'$. Since $T\not=T'$, there exists some formula $\theta\in T$ with $\theta\not\in T'$.
To show the desired conclusion, let $\phi$ be any arbitrary formula s.t. $ \Box\varphi \in T$, and we need to prove that $\varphi \in T' $.  From $\theta\in T$, we infer $(\varphi \vee \theta)\in T$; similarly, from $\Box\varphi \in T$, we infer $\Box (\varphi\vee \theta)\in T$.
Putting these together, we obtain $\refbox(\varphi\vee\theta)\in T$. By our assumption, this implies that $(\varphi\vee \theta)\in T'$, and since $\theta\not\in T'$, we conclude that $\varphi\in T'$, as desired.
\endproof
 \end{toappendix}

As a consequence of Lemma \ref {BoxStar}, we immediately get:

\begin{lemma}[Canonical $\refdia$-Witness Lemma]\label{starexistence}
For every formula $\varphi$ and theory $T\in \Omega$, we have that $\refdia\varphi\in T$ iff  there exists some theory $ T'\in \Omega$ s.t. $  T\crel  T' \ni\varphi.$
\end{lemma}

\smallskip

\par\noindent\textbf{Final Theories} Given a formula $\theta$, a theory $T\in \Omega$ is \emph{$\theta$-final} if we have: $\theta\in T$, and for all theories $S\in \Omega$, if $T\access S$ and $\theta\in S$ then $S\access T$ (hence $T\biaccess S$). Given a set $\Sigma$ of formulas, a theory $T\in \Omega$ is \emph{$\Sigma$-final} (or `final', for short) if it is $\theta$-final for some formula $\theta\in \Sigma$.

\smallskip
\par\noindent\textbf{Final Model}
Let $\Sigma$ be any set of formulas. The \emph{final model} is the canonical submodel\footnote{Any subset $X'\subseteq X$ of the set of worlds of a relational model
 $M=(X,\access,\|\bullet\|)$ determines a unique \emph{submodel}, obtained by taking: $X'$ as its set of worlds; the restriction of $\access$ to $X'$ as its accessibility relation; and the valuation given by $\|p\|\cap X'$.} determined by the set $\Omega^\Sigma:=\{T\in \Omega: T \mbox{ is $\Sigma$-final}\}$ of all final theories.

\smallskip

The final model may be infinite, but we can show that it has finite depth:

\begin{lemma}[Finite Depth Lemma]\label{WitnessingCluster}
The final model $\Omega^\Sigma$ has depth bounded by $|\Sigma|-1$. In other words: for every chain of $\Sigma$-final theories $T_0\srel T_1\srel \ldots T_n$, we have that $n\leq |\Sigma|-1$.
\end{lemma}

 \begin{toappendix}
\proof%[Proof of Lemma \ref{WitnessingCluster}]
Suppose, towards a contradiction, that $T_0\access T_1\access \ldots T_n$ is a strict chain of $\Sigma$-final theories of length $n\geq |\Sigma|$. Since all $T_i$ are $\Sigma$-final, there exist formulas $\theta_0, \ldots,\theta_n\in \Sigma$ s.t. $T_i$ is $\theta_i$-final (and hence $\theta_i\in T$) for all $i\leq n$. But this is a sequence of $n+1\geq |\Sigma|+1>|\Sigma|$ formulas in $\Sigma$, so some formula $\theta$ must be repeated. Let $\theta$ be such a repeating formula in the enumeration, and let $i$ and $j$ be indices such that $i<j$ and $\theta_i=\theta_j=\theta$.

So we have $T_i\access T_{i+1}\crel  T_j$, with both $T_i$ and $T_j$ being $\theta$-final, and so also $T_i\crel  T_j$. We have two cases: either $T_i\access T_j$ or $T_i=T_j$. We claim that in both cases we have $T_{i+1}\crel  T_i$. To show this, consider first the case $T_i\access T_j$. By $\theta$-finality we get $T_i\biaccess T_j$, hence $T_i\access T_{i+1}\crel  T_j\biaccess T_i$, and thus $T_i \access T_{i+1}\crel  T_i$, as desired. In the second case, we assume $T_i=T_j$, so we immediately obtain $T_{i+1}\crel  T_j=T_i$, as desired.

So we showed that we have $T_i\access T_{i+1}\crel  T_i$. There are again two cases: either $T_i\access T_{i+1}\access T_i$, or $T_i\access T_{i+1}=T_i$. In the first case, we immediately conclude that $T_i\biaccess T_{i+1}$, which contradicts the `strictness' of our chain. In the second case, we have $T_i\access T_{i+1} = T_i\access T_{i+1}=T_i$, so we again conclude that $T_i\biaccess T_{i+1}$, in contradiction with our `strictness' assumption.
\endproof
 \end{toappendix}

\medskip

In order to prove completeness with respect to the final model, we first need to show that every consistent formula belongs to some final theory. This is achieved by combining the Lindenbaum Lemma with the following.\david{There was no period here, wich seems odd to me. I'd normally use a period or a colon, unless the no-period thing was on purpose?}

\begin{lemma}[Final Lemma]\label{final}
If $\varphi\in T\in \Omega$, then there exists some $\varphi$-final theory $T^*\in \Omega$ such that $T\crel  T^*$ (and obviously, $\varphi\in T^*$, by finality).
\end{lemma}

 \begin{toappendix}

\proof%[Proof of Lemma \ref{final}]
We will use a well-known variant of Zorn's Lemma, stated for preorders: a preordered set $(\mathcal{S}, \leq)$ has a maximal element if every chain has an upper bound. (Here, being maximal in a preordered set means that there is no strictly larger element.)

Let $\varphi\in T\in \Omega$. Take $\mathcal{S}:=\{T'\in \Omega: T\crel  T'\ni\varphi\}$, with the relation $\crel $ as its preorder. Let $\mathcal{S'}\subseteq \mathcal{S}$ be a chain of theories in $\mathcal{S}$. To show that it has an upper bound, take the set
$$\Phi:=\{\varphi\} \cup \{ \refbox\theta : \refbox\theta\in T' \mbox{ for some } T'\in \mathcal{S'}\}$$
We show that \emph{$\Phi$ is consistent}: suppose this is not the case. Then there exists some \emph{finite} such inconsistent subset $\Phi'=\{\varphi\}\cup \{\refbox\theta_1, \ldots, \refbox\theta_n\}$, with $\refbox\theta_1\in T_1, \ldots, \refbox\theta_n\in T_n$ for some theories $T_1, T_2, \ldots, T_n\in \mathcal{S'}$. Since $\mathcal{S'}$ is a chain, we can assume that $T_1, T_2, \ldots T_{n-1}\crel  T_n$, and thus $\refbox\theta_1, \ldots, \refbox\theta_n\in T_n$. Since $T_n\in \mathcal{S}$, we also have $\varphi\in T_n$, so $\Phi'\subseteq T_n$, which contradicts the consistency of $T_n$.

Applying now Lindenbaum's Lemma, there exists some maximally consistent extension $S\in \Omega$ with $\Phi\subseteq S$. By construction (and using Lemma \ref{BoxStar}), we have $T'\crel  S$ for all $T'\in \mathcal{S'}$, so $S$ is an upper bound for the chain $\mathcal{S}$. Applying Zorn's lemma, we obtain a $\crel $-maximal element $T^*\in \mathcal{S}$.
In particular, this means that $\varphi\in T^*$ and $T\crel  T^*$, as desired. To prove that $T^*$ is $\varphi$-final, suppose that $T^*\access S\ni\varphi$; we have to show that $S\access T^*$. By the $\crel $-maximality of $T^*$, we must have $S\crel  T^*$, i.e. either $S\access T^*$ or $S= T^*$. If $S\access T^*$, then we are done. If $S=T^*$, then $S=T^*\access S=T^*$, so we get again $S\access T^*$, as desired.
\endproof

 \end{toappendix}

Using similar reasoning, we may establish an analogue of the $\Diamond$-Witness Lemma for final theories:

\begin{lemma}[Final $\Diamond$-Witness Lemma]\label{finalexistence}
For any theory $T\in \Omega$ and formula $\varphi$, we have that $\Diamond\varphi\in T$ iff there exists some $\varphi$-final theory $T'$ such that
$T\access T'$.
(Obviously, we have $\varphi\in T'$ in this case, by finality.)
\end{lemma}

 \begin{toappendix}

\proof%[Proof of Lemma \ref{finalexistence}]
The \emph{left-to-right implication}: by the Canonical $\Diamond$-Witness Lemma \ref{existence},  $\Diamond\varphi\in T$ implies the existence of some theory $S$ with $T\access S$ and $\varphi\in S$. By the Final Lemma \ref{final}, there exists some $\varphi$-final theory $S^*$ with $S \access S^*$ and $\varphi\in S^*$. If $T\access S^*$, then we can take $T':=S^*$ and we are done (since $S^*$ is $\varphi$-final and $T\access S\ni\varphi$, as desired). If $T\not\access S^*$, then from this and $T\access S\access S^*$ we get by weak transitivity that $T=S^*$, and so $T\access S\access S^*=T$. In this case, we can take $T':=S$. Indeed, since we already know that $T\access S\ni\varphi$,  to finish the proof we only need to check that $S$ is $\varphi$-final. For this, let $U\in \Omega$ be any theory with $S\access U \ni \varphi$; we need to show that $U \access S$.
  From $S^*=T\access S\access U$, we obtain by weak transitivity that either $U=S^*=T\access S$ (and we are done), or $S^*\access U \ni\varphi$. In the second case, by the $\varphi$-finality of $S^*$, we have $U\access S^*=T\access S$; by weak transitivity, we obtain either $U\access S$ (and we are done) or $U=S\access U=S$. So, in all cases, we concluded that $U\access S$, as desired.\david{Changed $W$ to $U$ to avoid the connotation with set of worlds.}

  The converse follows directly from the Canonical $\Diamond$-Witness Lemma \ref{existence}, as a special case.
  \endproof

\end{toappendix}

It will be useful to observe that $\theta$-final theories are closely related to $\refdia\theta$-final theories.

\begin{lemma}\label{Diamond}
Let $T\in\Omega$ be a $\theta$-final theory. Then:
\begin{enumerate}

\item $T$ is also $\refdia\theta$-final.

\item For every $S\in \Omega$ s.t. $T\crel  S$, we have $\refdia\theta\in S$ iff either $T=S$ or $T\biaccess S$.

\item All theories $S\in \Omega$ satisfying $T\biaccess S$ are $\refdia\theta$-final.

\end{enumerate}
\end{lemma}

 \begin{toappendix}
\proof%[Proof of Lemma \ref{Diamond}]
Assume $T$ is $\theta$-final. To show that it is also $\refdia\theta$-final, observe that we have $\refdia\theta\in T$ (since $\theta\Longrightarrow \refdia\theta$ is a theorem in our logic).
Second, let $S\in\Omega$ be s.t. $T\access S$ and $\refdia\theta\in S$, and we need to prove that $S\access T$. Since $\refdia\theta\in S$, we have either $\theta\in S$ or $\Diamond\theta\in S$. In the first case, from $T\access S\ni\theta$ and the fact that $T$ is $\theta$-final, we conclude that $S\access T$, as desired. In the second case, from $\Diamond\theta\in S$ we infer (by the Canonical $\Diamond$-Witness Lemma) that there exists $S'\in \Omega$, with $S\access S'\ni\theta$. Since $T\access S\access S'$, by weak transitivity we have either $T= S'$ or $T\access S'$. If $T=S'$, then we conclude $S\access S'=T$, and we are done. If $T\access S'$, then since $T$ is $\theta$-final and $\theta\in S'$, we get $S'\access T$. Thus we have $S\access S'\access T$, hence by weak transitivity we get that either $S\access T$ (and we are done) or $S=T$ (in which case $S=T\access S=T$, so we again obtain $S\access T$, as desired).

For the second claim of the Lemma: assuming $S\in \Omega$ s.t. $T\crel  S$ (i.e. $T=S$ or $T\access S$), we need to show that $\refdia \theta\in S$ holds iff either $T=S$ or $T\biaccess S$. The case $T=S$ is obvious. In the case $T\access S$, the left-to-right implication follows from the fact that $T$ is $\refdia\theta$-final. As for the converse: assuming $T\biaccess S$, and using the fact that
$\Diamond\theta\in T$, we obtain $\Diamond\refdia\theta\in S$ (by the Diamond Existence Lemma and $S\access T$), and so $\refdia\theta\in S$ (because $\Diamond\refdia\theta\Rightarrow \refdia\theta$ is a theorem in our axiomatic system).

For the third claim of the Lemma: assume that $S\in \Omega$ is s.t. $T\biaccess S$. Since this implies that $T\crel  S$, we are in the conditions of the second claim, and hence we can apply it to derive from $T\biaccess S$ that $\refdia\theta\in S$. To show finality, let $S'\in \Omega$ be s.t. $S\access S'$ and $\refdia\theta\in S'$; we need to prove that $S'\access S$. From $T\crel  S$ and $S\access S'$, we get $T\crel  S'$. From this and $\refdia\theta\in S'$, we obtain by the second claim that we have either $S'\biaccess T$ or $S'=T$. Both cases, combined with the fact that $T\crel S$, give us $S'\crel  S$. This means that we either have $S'\access S$ (as desired) or else $S'=S$ (in which case we have $S'=S\access S'=S$, hence we again get $S'\access S$, as desired).
\endproof
 \end{toappendix}
%

%\medskip

\par\noindent\textbf{Locality in the final model} For the rest of this section, whenever we talk about `locality', we refer to the final model $\Omega^\Sigma$. In particular, for $T\in \Omega^\Sigma$, we use the notations $T\uparrow:=\{S\in \Omega^\Sigma: T\rel S\}$,
$T\rupst:=\{S\in \Omega^\Sigma: T\crel S\}$, and whenever we write that a property holds locally ``above $T$", we mean that it holds above $T$ in $\Omega^\Sigma$.

\medskip

\par\noindent\textbf{Notation}. It is useful to introduce the notation
$$\widehat{\f}:=\{T\in \Omega^\Sigma: \f\in T\}$$
for all formulas $\f\in {\mathcal L}_\mu$. From the definition of the canonical valuation on (the canonical model, and hence on) the final submodel, it is obvious that we have $\|x\|_{\Omega^\Sigma}=\widehat{x}$, for all atoms $p\in P$. Our goal is to extend this observation to all sentences in $\Sigma$.

\medskip

\par\noindent\textbf{$\Sigma$-Bisimilarity in the Final Model}
We can apply the
concept $P$-bisimilarity $\simeq_P$ to the final model $\Omega^\Sigma$ (for any set of variables $P\subseteq\pvar$), and in fact the special case of $P_0$-bisimilarity $\simeq_{P_0}$ will be relevant for our proof. But it is convenient to introduce a stronger notion: a relation
$B\subseteq \Omega^\Sigma\times \Omega^\Sigma$ is a \emph{$\Sigma$-bisimulation} if it satisfies the same back-and-forth clauses as a usual $P$-bisimulation, but the Atomic Preservation clause is replaced by the requirement that: $(T, T')\in B$ implies $T\cap \Sigma=T'\cap \Sigma$. The relation ${\simeq_\Sigma} \subseteq \Omega^\Sigma\times \Omega^\Sigma$ of \emph{$\Sigma$-bisimilarity} is defined  as the largest $\Sigma$-bisimulation relation on $\Omega^\Sigma$.

It is easy to see that \emph{$\simeq_\Sigma$ is an equivalence relation on $\Omega^\Sigma$}, and that it is \emph{stronger than $P_0$-bisimilarity}: if $T\simeq_\Sigma T'$ then $T\simeq_{P_0} T'$. Using this and the above-mentioned well-known fact about invariance of $\mu$-calculus under standard bisimilarity, we immediately obtain the following:

\begin{lemma}\label{BisimInv}
All the formulas $\f\in {\mathcal L}_\mu^{P_0}$ are invariant under $\Sigma$-bisimilarity, i.e.
% their valuation $\|\varphi\|_{\Omega^\Sigma}$ in the final model is closed under $\Sigma$-bisimilarity:
if $T,T'\in \Omega^\Sigma $ satisfy $T\simeq_\Sigma T'$, then for all $\varphi\in {\mathcal L}_\mu^{P_0}$ we have $T\in \|\varphi\|$ iff $T'\in \|\varphi\|$.
\end{lemma}

\medskip

It is useful to introduce a more ``local" version of closure under bisimilarity.

\smallskip

\par\noindent\textbf{Closure under $\Sigma$ bisimilarity above a point} This is just a special case of the general notion of asserting a property locally: a set $X\subseteq \Omega^\Sigma$ is \emph{closed under $\Sigma$-bisimilarity above a theory $T\in \Omega^\Sigma$} if $X\cap T\rupst$ is closed under $\Sigma$-bisimilarity.

Of course, global closure implies local closure: if a set $X\subseteq \Omega^\Sigma$ is closed under $\Sigma$-bisimilarity, then it is also closed under $\Sigma$-bisimilarity above every $T\in \Omega^\Sigma$. Note also that: \emph{$X$ is closed under $\Sigma$-bisimilarity above $T$ iff $X\cap T\rupst$ is}.

\medskip

\par\noindent\textbf{Convention on global/local versions}
Sometimes we want to assert that both the global and the local version of a statement hold in the final model $\Omega^\Sigma$: e.g. if a certain premise holds, either globally or locally, then a certain conclusion holds, either globally or locally. To do this in a compact manner, we will state the global version, but adding in brackets the words ``above $T$", to include the local version as well. An example is the following result:

\begin{prop}\label{Bisim2}
If $\overline{X}=(X_1, \ldots, X_n)$ is a tuple of sets $X_i\subseteq \Omega^\Sigma$ that are closed under $\Sigma$-bisimilarity (above some $T\in \Omega^\Sigma$), and $\f=\f(\overline{x})\in \Sigma$ is a $\Sigma$-formula, then $\|\f(\overline{X})\|$ is also closed under $\Sigma$-bisimilarity (above $T$).
\end{prop}

 \begin{toappendix}
\proof%[Proof of Proposition \ref{Bisim2}]
We prove only the local version (since the proof of the global statement is just a simplification the local proof, obtained by omitting every mention of $T$). Let $Q=\{q_1,\ldots, q_n\}$ be a set of $n$ `fresh' propositional atoms (with $\pvar\cap Q=\varnothing$). We extend the valuation of the final model $\Omega^\Sigma$ to all the atoms in $\pvar\cup Q$, by putting $\|q_i\|=X_i\cap T\rupst$, for all $i\leq n$. Then, using the fact that all $X_i$ are invariant under $\Sigma$-bisimilarity above $T$ (together with the fact that $P_0\subseteq\Sigma$), it is easy to see that $\Sigma$-bisimilarity above $T$ implies $P_0\cup Q$-bisimilarity above $T$, i.e.: if $T',T''\in T\rupst$ are s.t. $T'\simeq_\Sigma T''$, then $T'\simeq_{P_0\cup Q} T''$. Putting this together with the fact that $\f(\overline{q})\in {\mathcal L}_\mu^{P_0\cup Q}$ is closed under $P_0\cup Q$-bisimilarity and using Lemma \ref{locality}, we conclude that $\|\f(\overline{X})\|\cap T\rupst=
\|\f(\overline{X\cap  T\rupst}) \| \cap T\rupst
=\|\f(\overline{\|q\|})\|=\|\f(\overline{q})\|$ is closed under $\Sigma$-bisimilarity above $T$, and hence $\|\f(\overline{X})\|$ is also closed under $\Sigma$-bisimilarity above $T$.\endproof
 \end{toappendix}

Next, we will use the following easy observation:

\begin{lemma}\label{Bisimilarity1}
If $T,T'\in \Omega^\Sigma$ are such that $T\biaccess T'$ and $T\cap \Sigma=T'\cap \Sigma$, then $T\simeq_\Sigma T'$.
\end{lemma}

 \begin{toappendix}
\proof%[Proof of Lemma \ref{Bisimilarity1}]
Take
$$B:=\{(T, T')\in \Omega^\Sigma \times \Omega^\Sigma: \, T\cap \Sigma=T'\cap \Sigma, \mbox{ and $T\biaccess^* T'$}\}$$
Clearly, to prove our lemma it is enough to show that $B$ is a $\Sigma$-bisimulation.

\medskip

For this, assume $(T,T')\in B$, and we have to check that $(T,T')$ satisfy the three clauses in the definition of a $P$-bisimulation:

\smallskip

\emph{``Atomic" preservation} is automatically ensured by the fact that $T\cap \Sigma=T'\cap \Sigma$.

\smallskip

For \emph{the forth condition}, let $S\in \Omega^\Sigma$ such that $T\access S$. We need to show that this, together with $(T,T')\in B$, implies the existence of some $S'\in  \Omega^\Sigma$ with $T'\access S'$ and $(S,S')\in B$:

If $T=T'$, we can take $S':=S$, and we are done, since $(S,S')=(S,S)\in B$. Otherwise, we have $T\biaccess T'$ and $T\access S$. From these, we infer that we have either $T'\access S$ or $T'=S$. In the first case, we can take again $S':=S$, and we are done. In the second case, $T'=S$ yields $T\biaccess T'=S$, hence we can take $S':=T$: we then have $T'\access T=S'$ and $(S, S')=(T', T)\in B$ (by the symmetry of $B$ and the fact that $(T,T')\in B$), as desired.

\smallskip

The \emph{back condition} follows from the satisfaction of the forth condition and the symmetry of $B$.
\endproof
 \end{toappendix}

\par\noindent\textbf{Notations: (sets of) $\Sigma$-bisimilarity classes}. It is convenient to introduce a notation for $\Sigma$-bisimilarity classes over the final model: for every final theory $T\in \Omega^\Sigma$, we put
$$T_\Sigma \, :=\, \{S\in \Omega^\Sigma: T\simeq_\Sigma S\}$$
for the $\Sigma$-bisimilarity class of $T$. For every \emph{set} $\mathcal{S}\subseteq \Omega^\Sigma$ of final theories, we put
$$\mathcal{S}_\Sigma  \, :=\, \{S_\Sigma: S\in \mathcal{S}\}$$
for the set of $\Sigma$-bisimilarity classes of theories in $\mathcal{S}$. In particular,
for the case of the set $\Omega^\Sigma$ of \emph{all} final theories, we simplify the notation, writing
$$\Omega_\Sigma\, :=\,  (\Omega^\Sigma)_\Sigma=\{S_\Sigma: S\in \Omega^\Sigma\}$$
for the set of $\Sigma$-bisimilarity classes of all $\Sigma$-final theories. Similarly,
for each number $n$, we put
\begin{align*}
\Omega^n_\Sigma \  &:= \  \{T\in \Omega^\Sigma: \dpt(T)\leq n\}_\Sigma \\
&= \ \{T_\Sigma: T\in \Omega^\Sigma \mbox{ with } \dpt(T)\leq n\}
\end{align*}
for the set of all $\Sigma$-bisimilarity classes of theories of depth no larger than $n$.
By the Finite Depth Lemma \ref{WitnessingCluster}, we have $\Omega_\Sigma=\Omega^N_\Sigma$ for some natural number $N$.

\begin{prop}\label{finitebisim}
There are only finitely many distinct bisimilarity classes in the final model $\Omega^\Sigma$.
\end{prop}

 \begin{toappendix}
\proof%[Proof of Proposition \ref{finitebisim}]
It is enough to show that, \emph{for each natural number $n$, the set $\Omega^n_\Sigma$ is finite} (since the desired conclusion will obviously follow from the above observation that the set of all final $\Sigma$-bisimilarity classes $\Omega_\Sigma$ coincides with $\Omega^N_\Sigma$ for some number $N$).

The finiteness of $|\Omega_\Sigma^n|$ for all $n$ follows immediately by induction from the following two claims:
\begin{enumerate}
\item $|\Omega^0_\Sigma|\leq 2^{|\Sigma|}\cdot 2^{2^{|\Sigma|}}$;
\item $|\Omega^n_\Sigma|\leq 2^{|\Sigma|}\cdot 2^{2^{|\Sigma|}}\cdot 2^{|\Omega^{n-1}_\Sigma|}$ for all $n>0$.
\end{enumerate}

To prove these two claims, note first that, for every final theory $T\in \Omega^\Sigma$, its bisimilarity class $T_\Sigma$ is uniquely determined by the pair $(T\cap\Sigma, T{\uparrow}_\Sigma)$, where $T{\uparrow}_\Sigma=\{S_\Sigma: T\access S\}$ is the set of $\Sigma$-bisimilarity classes of $T$'s successors.  We can split further this second component into two parts, depending on whether these bisimilarity classes are of the same depth as $T$ or of lower depth. In other words: for a final theory $T$ of depth $n$, its $\Sigma$-bisimilarity class $T_\Sigma$ is uniquely determined by the \emph{triplet}
$$(T\cap \Sigma, T{\uparrow}_\Sigma - \Omega^{n-1}_\Sigma, T{\uparrow}_\Sigma\cap \Omega^{n-1}_\Sigma),$$
where the third component is empty when $n=0$.

To count these triplets, note that the number of distinct possibilities for the first component of the triple $T\cap\Sigma\subseteq \Sigma$ is at most $2^{|\Sigma|}$. Further, since $T\access S$ and $\dpt(S)\geq n=\dpt(T)$ implies $T\biaccess S$, we have that $T{\uparrow}_\Sigma - \Omega^{n-1}_\Sigma\subseteq \{S_\Sigma: T\biaccess S\}$. But, by Lemma \ref{Bisimilarity1} (and the fact that $T\biaccess S, S'$ implies by weak transitivity that we have either $S=S'$ or $S\biaccess S'$), distinct elements $S_\Sigma\not=S'_\Sigma$ of this last set must have $S\cap\Sigma\not=S'\cap \Sigma$. So the set $\{S_\Sigma: T\biaccess S\}$ has at most $2^{|\Sigma|}$ elements, and thus the number of distinct possibilities for the second component of the triple is at most
$2^{2^{|\Sigma|}}$. Finally, for the third component, we have $T{\uparrow}_\Sigma\cap \Omega^{n-1}_\Sigma \subseteq \Omega^{n-1}_\Sigma$ for $n>0$ (and is empty for $n=0$), so the number of distinct possibilities for the third component is at most
$2^{|\Omega^{n-1}_\Sigma|}$ for $n>0$ (and is $=0$ for $n=0$). The above two claims immediately follow.
\endproof
 \end{toappendix}

\begin{cor}\label{finiterank}
For every fixed-point formula $\nu y.\f(y,\tuple{x})$ where the values of $\tuple x$ are closed under $\Sigma$-bisimilarity, the iterative process in Proposition \ref{Mono}(3) reaches its fixed point on the final model (above some $T\in \Omega^\Sigma$) at some \emph{finite} stage. \emph{More precisely}: for all tuples $\tuple{X}$ of subsets of $\Omega^\Sigma$ that are closed under $\Sigma$-bisimilarity (above some $T\in\Omega^\Sigma$), there exists some $N$ s.t. we have that
$$\|\nu y.\f(y,\tuple{X})\|=\bigcap_n \f_y^n(\tuple{X})=\f_y^N(\tuple{X})
\, \mbox{ holds (above $T$)},$$
where $\f_y^0(\tuple{X}):=\Omega^\Sigma$, $\f_y^{n+1}(\tuple{X})=\|\f( \f_y^n(\tuple{X}),\tuple{X})\|$ (and all the formulas are interpreted in the final model $\Omega^\Sigma$).
\end{cor}

 \begin{toappendix}
\proof%[Proof of Corollary \ref{finiterank}]
It is obvious that the sequence
$$\f_y^0(\tuple{X})\supseteq \f_y^1(\tuple{X})\supseteq \ldots \f_y^n(\tuple{X})
\supseteq \ldots$$
stabilizes, reaching the fixed point (above $T$) at the first stage $N$ s.t.
$\f_y^N(\tuple{X})= \f_y^{N+1}(\tuple{X})$ holds (above $T$), provided that such a finite number $N$ exists. To show that such an $N$ exists, suppose towards a contradiction that all the sets
$\f_y^n(\tuple{X}) \setminus \f_y^{n+1}(\tuple{X})$ are non-empty (above $T$). For every $n$, let $T_n\in \f_y^n(\tuple{X})\setminus \f_y^{n+1}(\tuple{X})$. From the definition of the sequence $\f_y^n(\tuple{X})$
and Proposition \ref{Bisim2}, it follows (by an easy induction) that all the sets $\f_y^n(\tuple{X})$
are closed under $\Sigma$-bisimilarity (above $T$). So, when $T_n$ (in $T\rupst$) is eliminated in the move from stage $n$ to stage $n+1$, the whole $\Sigma$-bisimilarity class of $T_n$ (above $T$) is also eliminated. But since there are only finitely many $\Sigma$-bisimilarity classes (above $T$) in $\Omega^\Sigma$, this elimination process cannot go forever. In fact, an upper bound for the stabilizing stage $N$ is given by the number of bisimilarity classes.\endproof
 \end{toappendix}

\begin{lemma}[Functional Truth Lemma]\label{Truth}
For every formula $\f=\f(\tuple{y})\in \Sigma$ in which the variables in the string $\tuple{y}=(y_1, \ldots, y_n)$ are free (or do not occur), every $\Sigma$-final theory $T\in\Omega^\Sigma$, and every tuple $\overline{\theta}=(\theta_1, \ldots, \theta_n)$ of formulas $\theta_i\in \mathcal{L}_\mu^{P_0}$ s.t. $\widehat{\theta_i}$ is closed under $\Sigma$-bisimilarity above $T$, we have:
\begin{description}
\item[(1)] $T\in \|\f(\overline{\widehat{\theta}})\|$ iff $T\in \widehat{\f(\tuple{\theta})}$;
\item[(2)] if  $\f=\f(\tuple{y})=\f(z, \tuple{y})$ is positive in $z$, then
for all natural numbers $n\in N$, we have:
\begin{itemize}
\item $\f_z^n(\overline{\widehat{\theta}})=\widehat{\f_z^n({\overline{\theta}})}$ holds above $T$;
\item moreover, $\widehat{\f_z^n({\overline{\theta}})}$ is closed under $\Sigma$-bisimilarity above $T$;
\end{itemize}
\end{description}
where: $\|\f\|$ is the interpretation of $\f$ in the final model $\Omega^\Sigma$; $\f_z^n(\overline{\widehat{\theta}})$ is an instance of the sequence of sets in Corollary \ref{finiterank} (i.e. it is recursively defined by putting $\f_z^0(\overline{\widehat{\theta}}):=\Omega^\Sigma$, $\f_z^{n+1}(\overline{\widehat{\theta}}):=
\|\f( \f_z^n(\overline{\widehat{\theta}}), \overline{\widehat{\theta}})\|$); and $\f_z^n({\overline{\theta}})$
is a sequence of formulas, recursively defined by putting $\f_z^0({\overline{\theta}}):=\top$,
 $\f_z^{n+1}({\overline{\theta}}):= \f(\f_z^n({\overline{\theta}}), \overline{\theta})$.
\end{lemma}

 \begin{toappendix}

\proof%[Proof of Lemma \ref{Truth}]
We prove both assertions (1) and (2) by \emph{double induction on the depth} $\dpt(T)$ of $T\in \Omega^\Sigma$ and \emph{on the subformula-complexity} of $\f$.

\medskip

\emph{Proof of assertion (1)}:

\smallskip

\emph{The base cases} $\f:=y_i$ and $\f:=x\in P_0$, as well as the \emph{Boolean cases}
$\neg \f$ and $\f\wedge \f'$, are trivial.

\medskip

\emph{Case $\Diamond\f$}. We have the sequence of equivalencies: $T\in \|\Diamond\f(\tuple{\widehat{\theta}})\|$ iff $\exists S\in \|\f (\tuple{\widehat{\theta}})\| \mbox{ s.t. }
 T\access S$ (by the semantic clause for $\Diamond$ in the final model) iff  $\exists S\in \widehat{\f(\tuple{\theta})} \mbox{ s.t. }
 T\access S$ (by the induction hypothesis for $\f$) iff $T\in \widehat{\Diamond\f(\tuple{\theta})}$ (by
the Final $\Diamond$-Witness Lemma \ref{finalexistence}).

\medskip

\emph{Case $\nu x.\f$ with $\f=\f(x,\tuple{y})$}. Since $T$ is $\Sigma$-final, there exists some $\rho\in \Sigma$ s.t. $T$ is $\rho$-final, and so (by Lemma \ref{Diamond}) $T$ is also $\refdia\rho$-final, and moreover $\refdia\rho$ locally defines $T$'s cluster above $T$. Also by Lemma \ref{Diamond}, \emph{all} theories $S\in \Omega$ s.t. $T\biaccess S$ are also $\refdia\rho$-final, hence they all belong to $\Omega^\Sigma$ (since $\refdia\rho$ is provably equivalent to some $\Sigma$-formula)\footnote{This follows from $\rho\in \Sigma$, together with the fact that $\Sigma$ is closed under the $\refdia$ operator up to logical equivalence.}. For each theory $S$ in the cluster of $T$ (i.e. s.t. either $T\biaccess S$ or $T=S$), we put
$$\chi_S \,\,\,\, := \,\, \, \, \bigwedge\{\psi: \psi\in S\cap\Sigma\}.$$
Note that, for any theory $T'\in \Omega$, we have $\chi_S\in T'$ iff $T'\cap\Sigma=S\cap \Sigma$.
Put
$$\chi \,\,\,\, := \,\, \, \, \bigvee\{\chi_S: S\in \|\nu x. \f(\overline{\widehat{\theta}})\|, \mbox{ and }T\biaccess^* S\}.$$
Take now the sentence
$$\eta \,\,\,\, := \,\, \, \, (\refdia\rho \wedge \chi)\vee (\refbox\neg\rho\wedge\nu x. \f(x,\tuple{\theta}))$$

\smallskip

\textbf{Claim 1:} $\|\nu x. \f(\overline{\widehat{\theta}})\| =\widehat{\eta}$ holds above $T$.

\smallskip

\emph{Proof of Claim 1}: Let $S\in T \rupst$, i.e. s.t. $T\access S$. We need to show that: $S\in
\|\nu x. \f(\overline{\widehat{\theta}})\|$ iff $\eta\in S$. For this, we distinguish two cases.

\emph{Case 1: $T\biaccess S$ or $T=S$}. By Lemma \ref{Diamond}, we have $\refdia\rho\in S$. Then the desired conclusion follows from the following sequence of equivalencies:
$\eta\in S$ iff $\chi\in S$ iff $\exists S'\in  \|\nu x. \f(\overline{\widehat{\theta}})\| \mbox{ s.t. }
(S'\biaccess T \mbox{ or } S'=T)\, \& \, S'\cap \Sigma=S\cap \Sigma$ iff $S\in \|\nu x. \f(\overline{\widehat{\theta}})\|$ (by Lemma \ref{Bisimilarity1}, Proposition \ref{Bisim2} and the assumption that
all $\theta_i$ are closed under $\Sigma$-bisimilarity above $T$).

\emph{Case 2: $T\centernot\biaccess S$ and $T\not=S$}, hence $T\srel S $, and thus $\dpt(S)<\dpt(T)$. By Lemma \ref{Diamond}, we have $\refdia\rho\not\in S$, so $\refbox\neg\rho\in S$. Once again, the desired conclusion follows from the sequence of equivalencies:
$\eta\in S$ iff $\nu x. \f (\bar{\theta}) \in S$ iff $S\in \widehat{\nu x. \f (\bar{\theta})}$ iff $S\in \|\nu x. \f(\overline{\widehat{\theta}})\|$ (by the induction hypothesis for theories $S\in\Omega^\Sigma$ with $\dpt(S)<\dpt(T)$).

\medskip

Given Claim 1, we can now prove:

\smallskip

\textbf{Claim 2.} $\widehat{\eta}$ is closed under $\Sigma$-bisimilarity above $T$.

\smallskip

\emph{Proof of Claim 2}: By Claim 1, we have $\widehat{\eta}\cap T\rupst=
\|\nu x. \f(\overline{\widehat{\theta}})\|\cap T\rupst$, and the right-hand side can be easily seen to be closed under $\Sigma$-bisimilarity above $T$ (using Proposition \ref{Bisim2} and the assumption that all $\widehat{\theta}_i$ are closed under $\Sigma$-bisimilarity above $T$). Hence, $\widehat{\eta}$ is also closed under $\Sigma$-bisimilarity above $T$.

\smallskip

\textbf{Claim 3.} $(\eta\Rightarrow \nu x. \f(x,\tuple{\theta}))\in T$.

\smallskip

\emph{Proof of Claim 3}: Suppose not. Then by Proposition \ref{Theorems}(4), we must have $\refbox(\eta\Rightarrow \f(\eta,\tuple{\theta}))\not\in T$. By the Canonical $\refdia$-Witness Lemma \ref{starexistence}, there exists $T'\in \Omega$ (not necessarily final!) such that $T\crel  T'$, $\eta\in T'$ and $\f(\eta,\tuple{\theta})\not\in T'$. Once again, we distinguish two cases.

\emph{Case 1: $\refdia\rho\in T'$}. From $T\crel  T'$ and the $\refdia\rho$-finality of $T$, we obtain that $T'$ is also $\refdia\rho$-final, hence $T'\in \Omega^\Sigma$ and thus $T'\in \widehat{\eta}\cap T\rupst$ (since $\eta\in T'\in \Omega^\Sigma$ and $T\crel  T'$). We have the following sequence of equalities:

$\widehat{\eta}\cap T\rupst = \|\nu x. \f(\overline{\widehat{\theta}})\|\cap T\rupst=
\|\f(\|\nu x. \f(\overline{\widehat{\theta}})\|, \overline{\widehat{\theta}})\| \cap T\rupst=
\|\f(\|\nu x. \f(\overline{\widehat{\theta}})\|\cap T\rupst, \overline{\widehat{\theta}})\|\cap T\rupst=
\|\f(\widehat{\eta}\cap T\rupst, \overline{\widehat{\theta}})\|\cap T\rupst=
\|\f(\widehat{\eta}, \overline{\widehat{\theta}})\|\cap T\rupst=
\widehat{\f(\eta, \tuple{\theta})}\cap T\rupst$

(where we used repeatedly Claim 1, Lemma \ref{locality}, the fact that $\|\nu x. \f(\overline{\widehat{\theta}})\|$ is a fixed point of $X\mapsto \|\f(X, \tuple{\widehat{\theta}})\|$, as well as the induction hypothesis for $\f$, combined with the fact that all $\widehat{\theta_i}$'s and $\widehat{\eta}$ are closed under $\Sigma$-bisimilarity above $T$). Using the above equalities, we get from $T'\in \widehat{\eta}\cap T\rupst$ to $T'\in \widehat{\f (\eta, \tuple{\theta})}$, which contradicts the above assumption that $\f(\eta,\tuple{\theta})\not\in T'$.

\emph{Case 2: $\refbox\neg \rho \in T'$}. From this, together with $\eta\in T'$, we obtain (using reasoning in the axiomatic system) that $(\refbox\neg \rho \wedge \nu x. \f(x,\tuple{\theta}))\in T'$.
Using the Fixed Point Axiom, we get $(\refbox\neg \rho \wedge \f(\nu x. \f(x,\tuple{\theta}),\tuple{\theta}))\in T'$. Applying Proposition \ref{Theorems}(2) (and the fact that $\f$ is positive in $x$), we infer that
$\f(\refbox\neg \rho \wedge \nu x. \f(x,\tuple{\theta}))\in T'$, then applying Proposition  \ref{Theorems}(3)
(as well as the fact that $(\refbox\neg \rho \wedge \nu x. \f(x,\tuple{\theta}))\Rightarrow \eta$ is provable in propositional logic, hence by Necessitation $\refbox((\refbox\neg \rho \wedge \nu x. \f(x,\tuple{\theta}))\Rightarrow \eta)$ is a theorem in our system), we obtain that $\f(\eta,\tuple{\theta})\in T'$, which again contradicts the above assumption that $\f(\eta,\tuple{\theta})\not\in T'$.

\medskip

Given the above three Claims, let us prove the case $\nu x.\f$. For the \emph{left-to-right direction}: assume that $T\in \|\nu x. \f(\overline{\widehat{\theta}})\| $. Then by Claim 1, we have $T\in \widehat{\eta}$, hence $\eta\in T$, and thus by Claim 3, we also have  $\nu x. \f(x,\tuple{\theta})\in T$, i.e. $T\in \widehat{\nu x. \f(x,\tuple{\theta})}$, as desired.

For the \emph{converse}: assume that $T\in \widehat{\nu x. \f(x,\tuple{\theta})}$. Using reasoning in the axiomatic system (making essential use of the Fixed Point Axiom), we see that $\nu x. \f(x,\overline{\theta})\Rightarrow \f_x^n(\overline{\theta})$ for all $n$, so we get  $T\in \widehat{\f_x^n(\overline{\theta})}$ for all $n$. By the inductive assertion (2) of our Lemma (for $\f$), we obtain that $T\in \f_x^n(\overline{\widehat{\theta}})$ for all $n$. So $T\in \bigcap_n \f_x^n(\overline{\widehat{\theta}})$. But, by Corollary \ref{finiterank} (and the fact that all $\widehat{\theta_i}$ are closed under $\Sigma$-bisimilarity above $T$), this last set is equal above $T$ with the greatest fixed point of the operator $X\mapsto \|\f(X, \tuple{\widehat{\theta}})\|$, i.e. with $\|\nu x. \f(\overline{\widehat{\theta}})\|$, and so we obtain the desired conclusion.

\medskip

\emph{Proof of assertion (2)}:
We prove assertion (2) of our Lemma for $\f$, using the fact that we proved assertion (1) of the Lemma for $\f$. The proof is by induction on $n$. For $n=0$: $\f_z^0(\overline{\widehat{\theta}})=W =\widehat{\f_z^0(\overline{\theta})}$, and  $W$ is obviously closed under $\Sigma$-bisimilarity. For the inductive step $n+1$, assume the assertion is true for $n$.
Then, for all theories $S\in T\rupst$, we have the following sequence of equivalencies:

$S\in \f_z^{n+1}(\overline{\widehat{\theta}})$
iff $S\in \|\f( \f_z^{n}(\overline{\widehat{\theta}}), \overline{\widehat{\theta}})\|$ iff
$S\in \|\f( \widehat{\f_z^n(\overline{\theta})}, \overline{\widehat{\theta}})\|$
 (by the inductive hypothesis (2) for $n$) iff
 $S\in \widehat{\f(\f_z^n(\overline{\theta}), \overline{\theta})}$ (by the induction hypothesis (1) for $\f$ and $S$, and using the closure of $\widehat{\theta}$ and of $\widehat{\f_z^n(\overline{\theta})}$
 under $\Sigma$-bisimilarity above $T$, by the inductive hypothesis (2) for $n$, and as a consequence their closure under $\Sigma$-bisimilarity above $S\in T\rupst$) iff
 $S\in \widehat{\f_z^{n+1}(\overline{\theta})}$.

That takes care of the first item in assertion (2) of our Lemma. As for the \emph{second item} of this assertion (closure of $\widehat{\f_z^{n}(\overline{\theta})}$ under $\Sigma$-bisimilarity above $T$): first, using Proposition \ref{Bisim2} and the recursive definition of $\f_z^n(\overline{\widehat{\theta}})$, an easy induction on $n$ shows that all $\f_z^n(\overline{\widehat{\theta}})$ are closed under $\Sigma$-bisimilarity above $T$; from this, together with the already proven first item of assertion (2), we conclude that all $\widehat{\f_z^{n}(\overline{\theta})}$ are also closed under $\Sigma$-bisimilarity above $T$.  \endproof
 \end{toappendix}

\begin{lemma}[Truth Lemma]\label{Simpletruth}
For every formula $\f\in \Sigma$, we have:
$$\|\f\|_{\Omega^\Sigma}=\widehat{\f}.$$
\end{lemma}

\proof For $\f=\f(y_1,\ldots, y_n)\in \Sigma$, apply the Functional Truth Lemma \ref{Truth} to formula $\theta_i:=y_i$.
\endproof

\medskip

%We can now finish our completeness proof.
\emph{Weak completeness for $\sf wK4$ frames} follows immediately from Lemma \ref{Simpletruth} (cf. Appendix). By Lemma \ref{Irref-drop}, this also applies to irreflexive $\sf wK4$ frames, hence to topological derivative spaces, and thus to arbitrary derivative spaces.

\begin{toappendix}

\medskip

\par\noindent
\textbf{\emph{ Proof of Completeness for $\sf wK4$ frames, topological derivative spaces, and general derivative spaces:
}}

Recall that we started with a consistent formula $\f_0$, and a set $\Sigma$ s.t. $\f_0\in \Sigma$ and $\Sigma$ is closed under negations and $\refdia$ up to logical equivalence. Take some $\Sigma$-final theory $T_0\in \Omega^\Sigma$ with $\f_0\in T_0$ (-such a theory exists by the Lindenbaum Lemma combined with the Final Lemma \ref{final}). Since $\f_0\in T_0\in
\Omega^\Sigma$, the above Truth Lemma \ref{Simpletruth} shows that $T_0\models \f_0$ holds in $\Omega^\Sigma$. Hence, our axiomatic system is complete for the class of weakly transitive relational models. By Lemma \ref{Irref-drop}, we can add irreflexivity: the logic is the same, so the system is also complete for the class of irreflexive and weakly transitive models. But, as already mentioned in Example \ref{AlexDerivative}, this class coincides with the class of Alexandroff topological derivative models. So the system is also complete for topological derivative models (and thus also for general derivative models).

\end{toappendix}

\bigskip

As for \emph{finite model property}, this can be shown by taking the quotient of $\Omega^\Sigma$ modulo $\Sigma$-bisimilarity:

\medskip

\par\noindent\textbf{Final Quotient} The \emph{final quotient} $(\Omega_\Sigma,
\access_\Sigma, \|\placeholder\|_\Sigma)$ is defined as the ``strongly extensional $\Sigma$-quotient" of the final model $\Omega^\Sigma$;  i.e. the set of worlds $\Omega_\Sigma$ consists of all equivalence classes $T_\Sigma=\{S\in \Omega^\Sigma: T\simeq_\Sigma S\}$, the accessibility relation is given by putting $T_\Sigma\access_\Sigma S_\Sigma $ if there are $T' \in T_\Sigma$, $S'\in S_\Sigma$ s.t.~$T\access S$, and the valuation is given by putting, for each $p\in \pvar$, $T_\Sigma\in \|p\|_\Sigma$ iff there is $T'\in T_\Sigma$ s.t.~$T'\in \|p\|$.
% Since atoms in $P_0$ formulas are preserved under $\Sigma$-bisimilarity, this valuation is well-defined on $\Omega^\Sigma$.

\begin{prop}[Finite Model Property]\label{FinBis}\
\begin{enumerate}
\item The final quotient $\Omega_\Sigma$ is finite (with an upper bound given by a computable function of $|\Sigma|$);
\item for every formula $\f\in \mathcal{L}_\mu^{P_0}$, we have that: $\f$ is true in the final model at some final theory $T\in \Omega^\Sigma$ iff $\f$ is true in the final quotient at the $\Sigma$-bisimilarity class $T_\Sigma$;
\item $\mu$-calculus has FMP (wrt relational, topological and derivative-space semantics).
\end{enumerate}
\end{prop}

 \begin{toappendix}
\proof%[Proof of Proposition \ref{FinBis}]
The finite bound follows immediately from Proposition \ref{finitebisim}.

The second part follows from the easily checked fact that the map $T\mapsto T_\Sigma$ is a functional $P_0$-bisimulation between the two models, and that all $\mu$-calculus sentences $\f\in \mathcal{L}_\mu^{P_0}$ are invariant under $P_0$-bisimulations.

To check part 3 (FMP) for weakly transitive models, it is enough to check that the relation $\access_\Sigma$ is weakly transitive in the final quotient $\Omega_\Sigma$. For this, suppose that $T_\Sigma, S_\Sigma, U_\Sigma\in \Omega_\Sigma$ are s.t. $T_\Sigma\access_\Sigma S_\Sigma\access_\Sigma U_\Sigma$. By the definition of $\access_\Sigma$, this means that we can assume $T\access S$ and $S'\access U$, for some theory $S'\simeq_\Sigma S$. Since $\Sigma$-bisimilarity is a bisimulation relation, this implies that there exists some $U'\simeq_\Sigma U$ s.t. $S\access U'$. But from $T\access S\access U'$, we obtained (by weak transitivity) that we have either $T=U'$, in which case $T_\Sigma=U'_\Sigma=U_\Sigma$, or else $T\access U'$, in which case $T_\Sigma \access_\Sigma U'_\Sigma=U_\Sigma$, as desired.

Using again Lemma \ref{Irref-drop} (and the fact that the copying construction in its proof preserves a model's finiteness), we get FMP for irreflexive and weakly transitive models, i.e. (by the equivalence in Example \ref{AlexDerivative}) for Alexandroff topological derivative models (and hence for arbitrary topological derivative models, as well as general derivative models).
\endproof
 \end{toappendix}
\medskip

\par\noindent\textbf{Axiomatization of natural sublanguages}
The proofs in this section can all be carried out within any natural sublanguage $\mathcal L$ of $\mathcal L_\mu$.
In particular, natural sublanguages are closed under all operations used to define $\Sigma$, and moreover the formulas considered in the proof are all built from elements of $\Sigma$ using substitution, Booleans or applications of modalities.
This finishes the proof of Theorem \ref{Completeness}, establishing completeness and FMP for all logics $\mu\text{-}\mathsf{wK4}^\mathcal L$.
\medskip

Next, we get similar results for many logics above $\mu\text{-}\sf wK4$.

\section{Generalization to cofinal subframe logics}\label{secSubframe}

%Subframe logics over $\sf K4$ were introduced by Fine in \cite{Fin85}, where it is shown using the selection method that every subframe logic has the FMP. Zakharyaschev \cite{Zak96} generalized
%subframe logics to a larger class of cofinal subframe logics and showed that every cofinal subframe logic has the fmp. Later,  \cite{BGJ11} extended the class of cofinal subframe logic from extension of $\sf K4$ to  extensions  of $\sf wK4$ and proved that every cofinal subframe logic over $\sf wK4$ has the fmp.
%{\bf Nick: the above could probably be removed since it is already mentioned in the intro.}

Our completeness and finite model property uses only a handful of properties of the logic $\sf wK4$, and can readily be extended to a wide class of related logics.
To be precise, we will now show that FMP holds for any canonical cofinal subframe logic above $\sf wK4$ enriched with fixed-points.

\begin{definition}\label{def: cof}$($\cite[Ch.~9]{CZ97}$)$
Let $(W, \rel )$ be a weakly transitive frame. A subset $X\subseteq W$ is called \emph{cofinal} if $ X{{\uparrow}} \subseteq  X{\rdnst}$. That is, for every $x\in W$, if there is $y\in X$ such that  $y \rel x$, then there is  $z\in X$ with $x \crel z$.
\end{definition}

%\begin{remark}{\em
%Note that a more standard definition of cofinality is to require  that $W\subseteq \rdnst X$. This definition clearly implies the other. Also note that if $X$ is cofinal in the sense of
%Definition~\ref{def: cof}, then $\rupst X$ is cofinal in the latter sense. }
%\end{remark}

Let $\Lambda$ be any normal modal logic over $\mathcal L_\ps$ that extends $\mathsf{wK4}$.
Recall that a Kripke frame $(W, \rel)$ is called a $\Lambda$-frame if it validates all the formulas in $\Lambda$, and that a modal logic $\Lambda$ is \emph{canonical} if the underlying frame of the canonical model for $\Lambda$ is a $\Lambda$-frame.
Every logic axiomatized by Sahlqvist formulas is canonical \cite{BdRV01}.
Recall also that a canonical logic $\Lambda$ is {\em cofinal subframe} if and only if for every $\Lambda$-frame $\mathcal F = (W, \rel)$ and every cofinal $U\subseteq X$, the restriction of $\mathcal F$ to $U$ is also a $\Lambda$-frame \cite{CZ97}.%Also a modal logic $\Lambda$ is \emph{$\mathcal{D}$-persistent} if for every descriptive frame of $\Lambda$, its underlying Kripke frame is also a $\Lambda$-frame. Note that every $\mathcal{D}$-persistent logic is canonical and every logic axiomatized by Sahlqvist formulas is $\mathcal{D}$-persistent (e.g., \cite[Theorem 10.31]{CZ97}). Recall also that a $\mathcal{D}$-persistent logic $\Lambda$ is cofinal subframe iff for every $\Lambda$-frame $(W, R)$, its every cofinal subframe, with a restricted relation, is also a $\Lambda$-frame \cite{CZ97}.\nick{I will check whether $\mathcal{D}$-persistence can be eliminated.}

Examples of canonical cofinal subframe logics above $\sf wK4$ are $\sf{wKT_0}$, $\sf{K4} $,  $\mathsf{K4D} = {\sf K4} + \Diamond \top$,  $\mathsf{K4.1} = \mathsf{K4} + \Box\Diamond p \to \Diamond \Box p$, $\mathsf{K4.2} = \mathsf{K4} + \Diamond \Box p \to \Box \Diamond p$, $\mathsf{K4.3} =  \mathsf{K4} + \Box(\Box^+ p\to q) \vee \Box (\Box^+q\to p)$, $\mathsf{S4} = \mathsf{K4} + \Box p \to p$, $\mathsf{S4.1} = \mathsf{S4} + \Box\Diamond p \to \Diamond \Box p$, $\mathsf{S4.2} = \mathsf{S4} + \Diamond \Box p \to \Box \Diamond p$, $\mathsf{S4.3} =  \mathsf{S4} + \Box(\Box p\to q) \vee \Box (\Box q\to p)$, $\mathsf{S5} = \mathsf{S4} + p\to \Box\Diamond p$,
%every extension of $\sf S4.3$
etc.~(see \cite[Chapter 9]{CZ97}).\david{I removed some definitions already mentioned previously.}

%\begin{definition}
%Let $\Lambda$ be a modal logic. We denote by $L^\mu$ the extension of $L$ by fixed point operators. That is, $L^\mu$ is %the least set of formulas in the modal language enriched with fixed point operators that contains all the formulas of $L$ and the Fixed Point axiom, and is closed under modus ponens, necessitation, uniform substitution and the Induction Rule.
%\end{definition}

%It is easy to see that a frame is a $\Lambda$-frame iff it is an $L^\mu$-frame.

We are ready to state Theorem \ref{Completeness} in full generality.

\begin{thm}\label{thmCofinal}
Let $\Lambda$ be a canonical cofinal subframe logic over $\sf wK4$, and $\mathcal L$ be a natural sublanguage of $\mathcal L_\mu$.
Then, $\mu\text{-}\Lambda^\mathcal L$ is sound for the class of $\Lambda$-frames, and complete for the class of finite $\Lambda$-frames.
\end{thm}

 \begin{toappendix}
\begin{proof}%[Proof of Theorem \ref{thmCofinal}]
We just follow the proof of the previous section. First, we note that by canonicity, the Kripke frame underlying the canonical model of $\mu\text{-}\Lambda$ is a $\Lambda$-frame.
This does require some checking, as canonicity only tells us that the canonical model for $\Lambda$ is a $\Lambda$-frame, but this can be done by observing that the canonical model of $\mu$-$\Lambda$ is a generated submodel of the canonical model of $\Lambda$.
We proceed as in the proof of weak completeness in the previous section with a small modification that if $\top$ does not belong to $\Sigma$, then we add it to $\Sigma$.
The fact that  $\top\in \Sigma$ ensures that
the final model $\Omega^\Sigma$ contains all final point of the canonical frame. Therefore, $\Omega^\Sigma$ is based on a cofinal subframe of the canonical frame. Hence, the underlying frame of $\Omega^\Sigma$ is a $\Lambda$-frame. Finally, as p-morphic images preserve the validity of modal $\mu$-formulas, the finite p-morphic image of $\Omega^\Sigma$ is a finite $\Lambda$-frame. Thus, every consistent $\mathcal L_\mu$-formula is satisfied in a model based on a
$\Lambda$-frame, and hence, on a $\mu$-$\Lambda$-frame, implying the FMP of $\mu$-$\Lambda$.
\end{proof}
 \end{toappendix}

There exist continuum many canonical cofinal subframe logics above $\sf K4$ (\cite[Theorem 11.28 and Exercise 11.14]{CZ97}). Hence the above theorem covers continuum many logics.\footnote{As far as we are aware, this is a first non-trivial example of a completeness result for modal fixed-point logics that covers so many logical systems.} Of course, only countable many of them have a recursively enumerable set of axioms: for those logics, decidability follows from Theorem \ref{thmCofinal}. Next, we single out some important ones.\footnote{Topological completeness of $\sf wK4$, $\sf{wKT_0}$ and $\sf{K4} $ has already been discussed in Section 3. We also recall that $\sf{S4.1}$ is the logic of spaces whose dense sets form a filter, that  $ \sf{S4.2}$ is the logic of  extremally disconnected spaces \cite[Sec.~2.6]{vBB07} and that $\sf{S4.3}$ is the logic of hereditarily extremally disconnected spaces \cite{BBL-BvM15}.}

\begin{corollary}\label{mulogics} The logics
$\mu\text{-}\sf{wKT_0}$,  $\mu\text{-}\sf{K4}$, $\mu\text{-}\sf{K4D}$, $\mu\text{-}\sf{K4.1}$, $\mu\text{-}\sf{K4.2}$, $\mu\text{-}\sf{K4.3}$, $\mu\text{-}\sf{S4}$, $\mu\text{-}\sf{S4.1}$,
$\mu\text{-}\sf{S4.2}$, $\mu\text{-}\sf{S4.3}$ have the FMP and are decidable.
\end{corollary}

\section{Completeness for $T_0$ and $T_D$ spaces}\label{secTopComp}

The simple world-duplication construction underlying the last step of the topological completeness proof in Section \ref{Completeness} does not work in the case of $T_0$ and $T_D$ spaces. So we will use d-morphisms to prove topological completeness for these cases.
In the process we give an alternative to the above-mentioned proof of completeness for arbitrary spaces, although this new proof has the disadvantage that it does not yield the finite model property in this setting.

For this it suffices, given a $\sf wK4$ frame $\mathcal F = (W,\rel)$, to construct a topological space $(X,\tau)$ and a d-morphism $\pi\colon X\to W$ as characterized by Lemma \ref{lemmDmorKripke}, in such a way that if $\mathcal F$ is a $\sf wK4T_0$ frame then $X$ will be $T_0$, and if $\mathcal F$ is a $\sf K4$ frame, then $X$ will be $T_D$.

\begin{definition}
Let $\mathcal F=(W,\rel)$ be a $\sf wK4$ frame. We build a topological space $(X,\tau) = (X_\mathcal F,\tau_\mathcal F)$ and a map $\pi\colon X\to W$ as follows.
Let $W^{\rm r}$ be the set of reflexive points of $W$ and $W^{\rm i}$ be the set of irreflexive points.
%Recall that we use the following notation: $\crel$ is the reflexive closure of $R$, $\rel $ is the antisymmetric part of $\crel$ (i.e.~$w\rel  v$ if $w \crel v$ but $v \not \crel w$) and $w\biaccess  v$ if $w\crel v$ and $v\crel w$.
Then, set
\[X = (W^{\rm r} \times \mathbb N) \cup (W^{\rm i} \times \{\omega\}),\]
and say that $U\subseteq X$ is open if whenever $(w,\alpha) \in U$, the following two properties are satisfied:
\begin{enumerate}

\item There is $n\in\mathbb N$ such that for all $(v,\beta)\in X$, $v \biaccess w $ and $\beta\geq n$ implies that $(v,\beta) \in U$.

\item If $(v,\beta)\in X$ and $w \srel v$ then $(v,\beta ) \in U$.

\end{enumerate}
Finally, set $\pi(w,\alpha) = w$.
\end{definition}

In other words, if an open set contains $(w,\alpha)$ then it contains all copies of $v$ whenever $w\crel v$, except possibly for cofinitely many in the case that $v \biaccess w$.

\begin{lemma}\label{lemmIsDmor}
If $\mathcal F = (W,\rel)$ is any $\sf wK4$ frame then $\tau_\mathcal F$ is a topology on $X_\mathcal F$ and $\pi\colon X_\mathcal F \to W$ is a d-morphism.
\end{lemma}

 \begin{toappendix}
\begin{proof}%[Proof of Lemma \ref{lemmIsDmor}]
Let $(X,\tau) = (X_\mathcal F, \tau_\mathcal F) $.
We omit the proof that $\tau $ is a topology, which proceeds by routine verification.
To see that $\pi$ is a d-morphism, we appeal to Lemma \ref{lemmDmorKripke}.
Let $(w,\alpha)\in X$ and note that $w = \pi(w,\alpha)$.
Define
\[O = \{(v,\beta): w\crel v \text{ and }\beta\geq 0\}.\]
It should be clear that $O$ is a neighborhood of $(w,\alpha)$.
Moreover, if $(v,\beta )\in O\setminus \{(w,\alpha)\}$, then $w\crel v$.
It follows that $w\rel  v$, except in the case where $w=v$ and $w$ is irreflexive.
But then $\alpha=\beta=\omega$, so that $(v,\beta ) = (w,\alpha)$, contradicting $(v,\beta )\in O\setminus \{(w,\alpha)\}$.
Hence $O\setminus \{(w,\alpha)\} \subseteq \pi^{-1}(w\upst)$, as needed.

Now let $U$ be any neighborhood of $(w,\alpha)$ and $v\in w\upst$; we must show that $v\in \pi(U \setminus \{(w,\alpha)\} )$.
Then, there is $n\in \mathbb N$ such that $\beta > n$ and $(v,\beta) \in X$ implies that $(v,\beta) \in U$.
If $v$ is reflexive, choose any $\beta \neq\alpha $ such that $n<\beta<\omega$.
Then, $(v,\beta) \in U \setminus \{(w,\alpha)\}$ and $\pi(v,\beta) = v$, as required.
If $v$ is irreflexive, then from $w\rel  v$ we obtain $v\neq w$. Thus $(v,\omega) \in U\setminus \{(w,\alpha) \}$ and $\pi(v,\omega) = v$.

Finally, $\pi$ is surjective since $\pi(w,0) = w$ if $w$ is reflexive and $\pi(w,\omega) = w$ if $w$ is irreflexive.
\end{proof}
 \end{toappendix}

\begin{lemma}\label{lemmIsLambdaSpace}
If $\mathcal F = (W,\rel)$ is any $\sf wK4$ frame then:
\begin{enumerate}

\item If $\mathcal F$ is a $\sf wK4T_0$ frame then $X_\mathcal F$ is $T_0$.

\item If $\mathcal F$ is a $\sf K4 $ frame then $X_\mathcal F$ is $T_D$.

\end{enumerate}
\end{lemma}

 \begin{toappendix}

\begin{proof}%[Proof of Lemma \ref{lemmIsLambdaSpace}]
Let $X =X_\mathcal F$.
First assume that $\mathcal F$ is a $\sf wK4T_0$ frame and let $(w,\alpha)\neq (v,\beta) \in X$.
If $w\centernot\crel v$ then $\{(u,\gamma)\in X: w\crel u\}$ is a neighborhood of $(w,\alpha)$ which is not a neighborhood of $(v,\beta)$.
The case where $v\centernot\crel w$ is symmetric, so we may assume $w\biaccess^* v$.
If $\beta<\omega$ then
\[U = \{(u,\gamma)\in X: w\crel u  \} \setminus \{(v,\beta )\}\]
is a neighborhood of $(w,\alpha)$ not containing $(v,\beta)$.
The case where $\alpha<\omega$ is symmetric.
So we are left with the case where $\alpha=\beta=\omega$.
Since $(w,\alpha)\neq (v,\beta)$, it follows that $w\neq v$.
Since $w\biaccess v$ and $\mathcal F$ is a $\sf wK4T_0$ frame, we cannot have that both $w,v$ are irreflexive; but if $w$ is reflexive then $\alpha<\omega$, contrary to our assumption, and similarly if $v$ is reflexive then $\beta<\omega$.
We conclude that the case $\alpha=\beta=\omega$ is impossible.

Now assume that $\mathcal F$ is a $\sf K4$ frame and let $(w,\alpha)\in X$; we must find open $U$ and closed $F$ such that $U\cap F = \{(w,\alpha)\}$.
Let
\begin{align*}
U&=\{(v,\beta) \in X: w \crel v\}\\
F&=\{(v,\beta) \in X: v\rel  w\} \cup \{(w,\alpha)\}.\\
\end{align*}
It should be clear that $U\cap F = \{(w,\alpha)\}$ and that $U$ is open, so we check only that $F$ is closed; that is, that the complement of $F$ is open.
So, define $O:= X\setminus F$.
Let $(v,\beta) \in O$.
If $(u,\gamma)\in X$ is such that $v\rel  u$, then we cannot have $u\crel w$ by transitivity, since this would lead to $v\rel  w$ and $(v,\beta)\in F$.
Thus $u\centernot\rel  w $ and $u\neq w$, yielding $(u,\gamma) \not\in F$ regardless of $\gamma$.
Next we check that there is $n$ so that if $\gamma>n$, $u\biaccess^*  v$ and $(u,\gamma ) \in X$, it follows that $(u,\gamma) \in O$.
If $v\centernot\biaccess^*  w$, then we may set $n=0$.
For then, $u\biaccess^*  v$ and $v\centernot\rel  w$ yield $u\centernot\rel  w$ and $u\neq w$, so that $(u,\gamma)\in O$ regardless of $\gamma$.
If $v\biaccess^*  w$, we claim that $w$ is reflexive.
If not, then $v\biaccess^*  w$ yields $v=w$ since $\mathcal F$ is a $\sf K4$ frame, and the definition of $X$ yields $\alpha=\beta=\omega$, so that $(v,\beta) = (w,\alpha)$ and $(v,\beta) \in O$ is impossible.
Thus $w$ is reflexive, so that $\alpha<\omega$. But then, $u\biaccess^*  v$ and $\gamma>\alpha$ yield $(u,\gamma) \in O$, so we may set $n=\alpha$.
\end{proof}
 \end{toappendix}

\medskip

We can now proceed to prove topological completeness for $T_0$ and $T_D$ spaces. In fact, the proof also works for the  $\sf wK4$ case of arbitrary spaces (but unlike the proof in the previous section it does \emph{not} give us finite model property)\footnote{On the other hand, neither $\sf wK4T_0$ nor $\sf  K4$ have the finite topological model property, so the next result cannot be improved upon.}:

\begin{thm}\label{topocompleteness}
\begin{enumerate}

\item $\mu$-$\sf wK4$ is sound and complete for the class of all topological spaces.

\item $\mu$-$\sf wK4T_0$ is sound and complete for the class of all $T_0$ topological spaces.

\item $\mu$-$\sf  K4$ is sound and complete for the class of all $T_D$ topological spaces.

\end{enumerate}
\end{thm}

 \begin{toappendix}

\begin{proof}%[Proof of Theorem \ref{topocompleteness}]
Let $\Lambda$ be any of the logics
$\mu\text{-}\mathsf{wK4}$,  $\mu\text{-}\mathsf{wK4T_0}$ or $\mu\text{-}\mathsf{K4}$.
Soundness of $\Lambda$ for its class of spaces follows from Lemma \ref{Soundness} and the fact that each fixed point-free fragment is sound for the respective class of spaces (see Section \ref{secMu}).
Since $\Lambda$ is a canonical subframe logic, by Kripke completeness, if $\varphi$ is not derivable then it is falsifiable on some $\Lambda$-frame $\mathcal F = (W,\rel)$.
Then, by Lemma \ref{lemmIsLambdaSpace}, $(X_\mathcal F,\tau_\mathcal F)$ is a $\Lambda$-space and $\mathcal F$ is a d-morphic image of $ X_\mathcal F$, so $\varphi$ is also falsifiable on $(X_\mathcal F,\tau_\mathcal F)$, as needed.
\end{proof}

\end{toappendix}

\section{Conclusion and comparison with other work}\label{Conclusion}

In this paper, we have studied the $\mu$-calculus over arbitrary topological spaces, as well as some natural subclasses, and obtained a general soundness and completeness result for the standard axiomatization.

\medskip

Our results are novel for several reasons. First, in the setting of Kripke semantics, neither completeness nor the FMP for weakly transitive frames were known, nor do they follow immediately from known results.
Moreover, our completeness proof appears to be the first such result for a variant of $\mu$-calculus that simultaneously applies to infinitely many logics and their respective classes of frames.

From the topological perspective, neither completeness nor decidability for non-$T_D$ spaces were known, nor they follow from known results. Unlike the transitive/$T_D$ case, our logics do not embed into standard $\mu$-calculus, or any of its known decidable extensions. This is in sharp contrast to the $T_D$/transitive case, where FMP and decidability follow via a simple encoding into standard $\mu$-calculus.\footnote{As already mentioned, the transitive closure of a relation can be encoded in $\mu$-calculus (and thus the decidability of $\mu$-calculus over transitive frames follows immediately from Kozen's result on the decidability of $\mu$-calculus over arbitrary frames). In contrast, the weakly-transitive closure of a relation does \emph{not} seem to be definable in $\mu$-calculus, and not even in its recent hybrid extension  \cite{SV01}. Weakly-transitive closure is definable only if one adds the binding operator from hybrid logic. But this increases the complexity of hybrid $\mu$-calculus, and the resulting logic is no longer known to have FMP (or to even be decidable).} Moreover, we showed that the tangled derivative is not expressively complete over the class of all topological (or even $T_0$) spaces, so we had to give a completeness proof that applies to the full language of $\mu$-calculus.

But note that even on $T_D$ spaces, our proof is the first to directly establish completeness over such spaces of a Kozen-type axiomatization for full $\mu$-calculus (rather than for some semantically equivalent modal logic). Prior work on $T_D$ spaces, mainly by Goldblatt and Hodkinson \cite{GH18}, had focused only on the tangled fragment.
Though this fragment is known to be co-expressive with $\mu$-calculus over $T_D$ spaces (and transitive frames), completeness for the full $\mu$-calculus over these spaces only follows if we combine the results in \cite{GH18} with Walukiewicz's proof of Kripke completeness for $\mu$-calculus. In contrast, our proof of completeness is self-contained (for both the $T_D$ and the non-$T_D$ case), taking advantage of the weak transitivity to give a streamlined proof tailored for the topological setting.

Furthermore, our results are based on an innovative use of the proof techniques using final submodels (due to Fine and Zakharyaschev).
This method has not been applied previously in a setting with fixed points, and provides a novel, general and relatively simple approach to dealing with fixed point logics over $\sf wK4$ frames (for which the filtration method, used in \cite{GH18} and elsewhere in the study of fixed point logics, does not seem to work).
In fact, even for the much easier case of topological \emph{closure spaces}, our method provides a simpler and more uniform way to reprove existing results: while Goldblatt and Hodkinson \cite{GH17} had to do a lot of work to show the FMP for $\sf S4$-tangle logic (and thus also for the semantically equivalent $\mu\text{-}\sf{S4}$), in Corollary \ref{mulogics} we get this result essentially for free from our general methods.

\smallskip

There are many open questions left within the context of topological fixed point logics.
The problem of finding a simple but expressively complete fragment of the $\mu$-calculus over $\sf wK4$, in analogy to the tangled fragment for logics over $\sf K4$, remains open. But we conjecture that topological $\mu$-calculus does indeed collapse to a simpler natural fragment, possibly the alternation-free fragment, with a proof along the lines of the similar argument for transitive frames in \cite{Dagostino}.
Anticipating such a development, we have set up our main completeness result in a modular fashion so that, if such a fragment ${\mathcal L}$ is ever found, the completeness for its natural axiomatization
$\mu\text{-}\mathsf{wK4}^{\mathcal L}$ will follow immediately from Theorem \ref{Completeness}.

\smallskip

Another line of inquiry that we leave open here is the problem of extending our methods to classes of spaces which enjoy topologically natural properties that do not correspond to any cofinal subframe logic.
The prime example here is that of {\em connected spaces,} whose modal logic is well understood in presence of the universal modality \cite{Sheh99}.
We believe that our methods can be extended to such settings, but some non-trivial modifications would be required.



\bibliographystyle{IEEEtran}

\bibliography{LICS}

\end{document}